\RequirePackage{fixltx2e}
\documentclass[aps,pre,onecolumn,showpacs,superscriptaddress,10pt]{revtex4-1}

\usepackage[normalem]{ulem}
\usepackage{xcolor}

\usepackage{bm}
\usepackage{graphicx}
\usepackage{latexsym}
\usepackage{amsmath}
\usepackage{hyperref}
\usepackage{pxfonts}
\usepackage{cancel}
\usepackage{bbding}
\usepackage{subfig}
\usepackage{float}
\usepackage{subfloat}
\usepackage{caption}
\usepackage[section]{placeins}
\usepackage{csquotes}
\DeclareGraphicsExtensions{.pdf,.png,.jpg}

\begin{document}

\title{\Large{Non-linear Ion-Wake Excitation by the \\ Time-Asymmetric Electron Wakefields of Intense Energy Sources \\ with applications to the Crunch-in regime}}

\author{Aakash A. Sahai} 
\email{aakash.sahai@gmail.com}
\affiliation{Department of Physics, Blackett Laboratory and John Adams Institute for Accelerator Sciences, Imperial College London, London, SW7 2AZ, UK \\ \& Department of Electrical Engineering, Duke university, Durham, NC 27708, USA}

\begin{abstract}
A model for the excitation of a non-linear ion-wake mode by a train of plasma electron oscillations in the non-linear time-asymmetric regime is developed using analytical theory and particle-in-cell based computational solutions. The ion-wake is shown to be a driven non-linear ion-acoustic wave in the form of a cylindrical ion-soliton. The near-void and radially-outwards propagating ion-wake channel of a few plasma skin-depth radius, is explored for application to ``Crunch-in" regime of positron acceleration. The coupling from the electron wakefield mode to the ion-mode dictates the long-term evolution of the plasma and the time for its relaxation back to an equilibrium, limiting the repetition-rate of a plasma accelerator. Using an analytical model it is shown that it is the time asymmetric phases of the oscillating radial electric fields of the nearly-stationary electron bubble that excite time-averaged inertial ion motion radially. The electron compression in the back of the bubble sucks-in the ions whereas the space-charge within the bubble cavity expels them, driving a cylindrical ion-soliton structure with on-axis and bubble-edge density-spikes. Once formed, the channel-edge density-spike is sustained over the length of the plasma and driven radially outwards by the thermal pressure of the wake energy in electrons. Its channel-like structure is independent of the energy-source, electromagnetic wave or particle beam, driving the bubble electron wake. Particle-In-Cell simulations are used to study the ion-wake soliton structure, its driven propagation and its use for positron acceleration in the ``Crunch-in" regime.

\end{abstract}

\maketitle 

\section{Introduction}

\begin{figure}[ht!]
	\begin{center}
   	\includegraphics[width=5in]{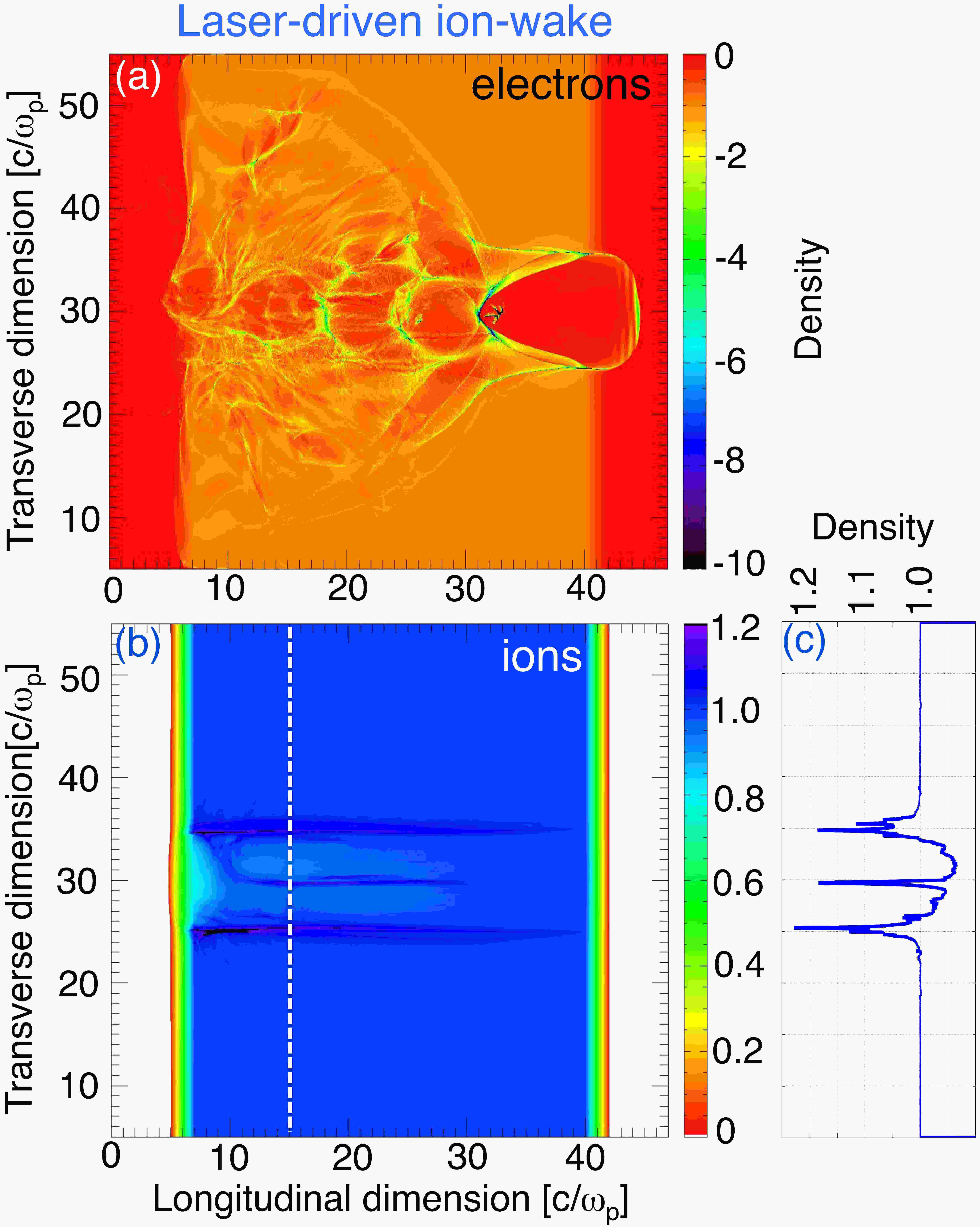}
	\end{center}
\caption{ \small {\bf Laser driven non-linear ion-wake at early time (t = 46$\omega_{pe}^{-1}$ = 0.17 $f_{pi}^{-1}$, where $f_{pi}$ is the plasma ion frequency) in $m_{i}=m_p=1836 ~ m_e$ plasma}. (a) Electron bubble wakefields in cartesian coordinates (fixed-box) with $\frac{\omega_0}{\omega_{pe}}=10$ driven by a matched laser pulse (vector potential $a_0=4$ and frequency $\omega_0$) with $R_B \simeq 4\frac{c}{\omega_{pe}}$. (b) Non-linear ion-wake in the form of a cylindrical ion-soliton of radius $\simeq 4\frac{c}{\omega_{pe}}$ excited behind the bubble electron wake in a proton plasma. (c) Transverse ion-density profile at $z=15~c/\omega_{pe}$.  Notice that the ion density perturbation in this excitation phase is still building up and is a fraction of the background ion density, $\frac{\delta n_i}{n_0} < 1$.}
\label{fig1:ion-wake-laser}
\end{figure}

Plasma ions are generally assumed to be stationary in the theory of ultra-relativistic non-linear plasma electron waves \cite{Akhiezer-Polovin}. Such electron waves are regularly excited as wakefields of high-intensity energy sources such as an ultra-short laser or particle beams and have proved to be promising for accelerating and transporting beams with unprecedented field strengths \cite{wakefield}\cite{cavitation-laser}\cite{cavitation-beam}\cite{Pukhov-laser-bubble}\cite{Lu-bubble-regime}. Important exceptions to the fundamental assumption of stationary ions occur as the intensities of the energy sources become high enough to lead to significant ion motion within a period of the electron wave. Ion motion also invariably becomes important over several periods of the electron wake train further behind the driver as the energy left-over in electron oscillation modes couples to the ion modes.
 
The motion of ions has significant implications for plasma acceleration as high-intensity conditions occur when the drive beam (or an accelerated witness beam) has fields that lead to ion trajectories that are a considerable fraction of the electron oscillation trajectory \cite{ion-motion-intense-beam}\cite{ion-motion-beam-emittance}. These conditions are predicted to arise in the final-stage of ultra-low emittance future plasma-based collider designs at the TeV energy scale. The subject of this paper however is the ion motion at longer timescales, understanding the long-term ion behavior is important to determine the state of the plasma for succeeding bunches in a high repetition rate future plasma-based collider \cite{long-term-wake}\cite{hot-plasma-wake}. The work presented here shows that a long-lived ion-mode is leftover in the plasma, establishing an upper-limit on the repetition-rate of the plasma-based accelerators.

In this paper, using theoretical analysis and computational modeling, the excitation of a nonlinear ion-wake in the trail of a non-linear ``bubble" plasma electron, is shown. The electron wake may be driven by either an intense laser or particle beam energy source \cite{cavitation-laser-expt}\cite{cavitation-beam-expt}. We show that the time asymmetry of the focusing fields of the bubble leads to the excitation of non-linear ion-acoustic modes in the form of a cylindrical ion-soliton. 

The application of the non-linear ion-wake for plasma-based accelerators in ``crunch-in" wakefield regime is explored. The ``crunch-in" regime of plasma wakefields in an ideal hollow-channel was introduced in \cite{positron-IPAC-2015}\cite{crunch-in-regime} and Ch.8 of \cite{Sahai-dissertation}. In this wakefield regime, it was shown that hollow-channel is driven by energy-sources such that the channel-wall electrons collapse to the axis, driving strong wakefields of the order of channel-wall cold-plasma wave-breaking fields. Importantly, it was also shown that the focusing fields excited in this regime have a linear radial dependence of magnitude (with direction favorable for positron transport) and are of the order of accelerating fields \cite{positron-IPAC-2015} (Fig.2 there-in). The excitation of strong focusing fields in this regime is completely opposite to the conventionally established conclusion that relativistic particles have zero focusing fields in hollow-channels \cite{Katsouleas-PRL-1998}. This regime is enabled by the ion-wake channel because it is shown to have an initial radius close to the electron wake transverse size while uniquely its length is as long as the acceleration length. Here we show that the ion-wake channel-wall electrons collapse towards the energy-propagation axis resulting in a non-linear on-axis electron density compression many times the near-void background density. The optimal compression is shown to be only possible if the driving beam properties are matched to the channel radius \cite{positron-IPAC-2015}, a strong dependence on the excitation which is a signature of non-linearity. The choice of appropriate channel radius is enabled by launching the driver at an appropriate time, resulting in excitation at an appropriate channel radius during the expansion of the ion-wake channel.

The ion-wake model shows two distinct phases of the non-linear ion-wake: inertial and thermalized phase. In the inertial or excitation phase the time-asymmetry between the attractive and repulsive radial fields of the bubble on the ions excites them into a soliton-like structure. We show that in this phase the inertial response of the ion rings is dictated by an equilibrium or separatrix radius. The ion rings located within this radius collapse towards the axis whereas rings outside are driven outwards. The outward propagating rings are only driven up to the bubble radius beyond which the force of the bubble radial fields rapidly falls off, resulting in the accumulation of the ion rings at the bubble radius.

At later times, the non-linear radial electron oscillations undergo phase mixing \cite{phase-mixing-longitudinal} leading to coherent electron motion becoming thermalized. The thermalized phase is shown to be a driven non-linear ion-acoustic wave in the form of a cylindrical ion-acoustic soliton. Its characteristics are similar to the solutions of the cylindrical Korteweg - de Vries equation (cKdV) \cite{Maxon-cyl-soliton}\cite{cyl-soliton-observation}\cite{Berezin-Karpman-1964}\cite{KdV-non-linear-ion-waves}. However, the ion-wake shown here differs from a cylindrical-KdV soliton in several aspects: (a) The bubble wake electron oscillations do not thermalize into an isothermal plasma, so the ion-wake soliton is driven (or forced) by the electron temperature gradient; (b) the ion-wake soliton breaks up into N-solitons as it evolves and (c) at early times there is an ion-density spike on the axis which collapses at a later time. The soliton propagates radially outwards leaving behind a flat residue resulting in a near-void ion-wake channel.    

\begin{figure}[ht!]
	\begin{center}
   	\includegraphics[width=5in]{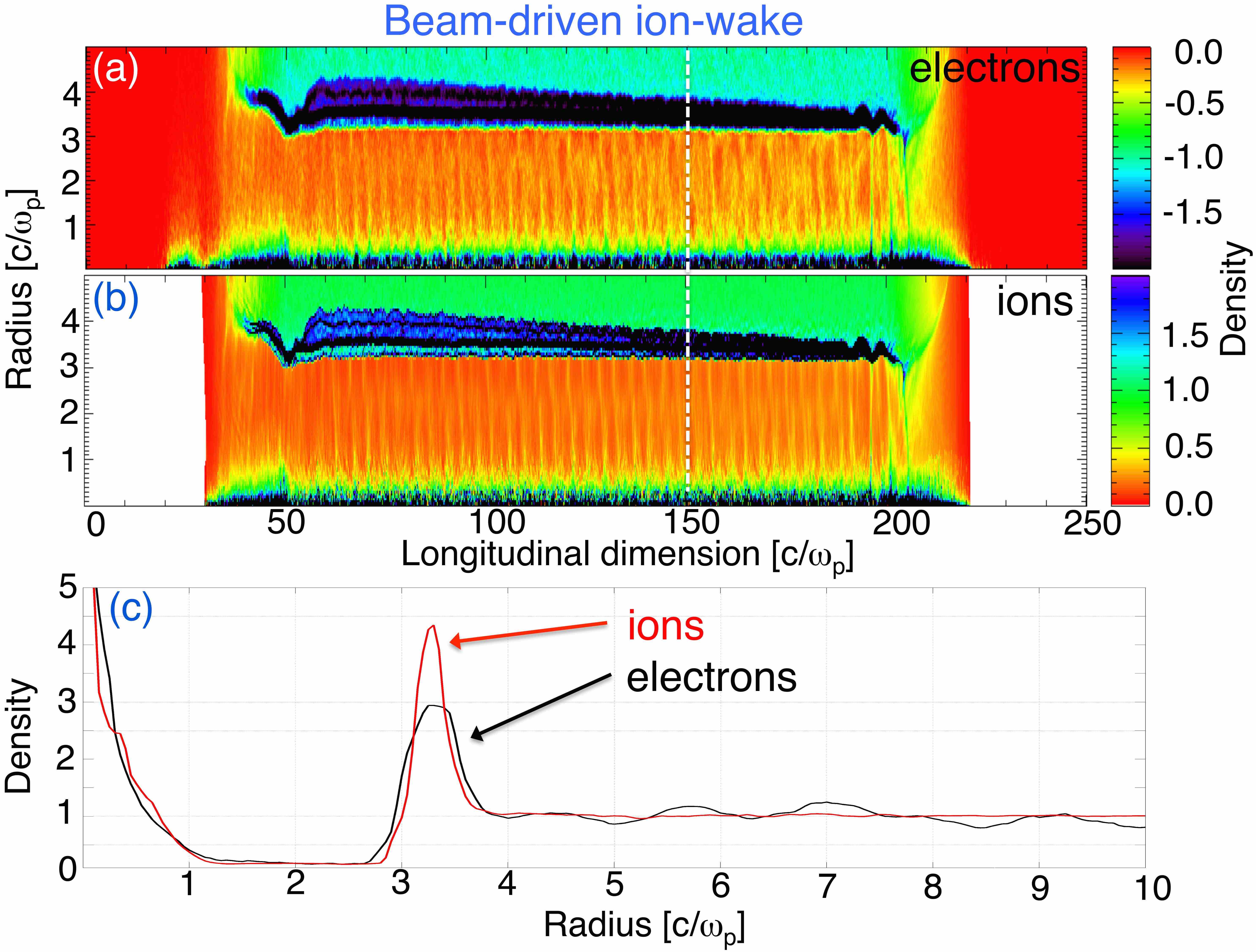}
	\end{center}
\caption{ \small {\bf Electron beam-driven non-linear ion-wake at late time (t = 460 $\omega_{pe}^{-1} = 1.7 f_{pi}^{-1}$) in $m_{i}=m_p=1836 ~ m_e$ plasma}. (a) Beam-driven ion-wake electron density in cylindrical coordinates (fixed-box). The beam parameters are $n_b=5n_0$, $\sigma_r=0.5c/\omega_{pe}$, $\sigma_z = 1.5 c/\omega_{pe}$, $\gamma_b=38,000$, these beam-plasma parameters are quite similar to \cite{cavitation-beam}. (b) Corresponding ion density in cylindrical coordinates (fixed-box). Note the N-soliton formation in the ion-density, $50c/\omega_{pe} \leq z \leq 100 c/\omega_{pe}$. The later times in the time-evolution of the ion-wake is also inferred from density snapshots farther behind the beam. (c) Radial electron and ion density profile at $z=150~c/\omega_{pe}$. A full movie of radial electron and ion density dynamics is presented in supplementary material \cite{supplementary-movie}. }
\label{fig2:ion-wake-beam}
\end{figure}

Representative PIC simulation results in Fig.\ref{fig1:ion-wake-laser} and Fig.\ref{fig2:ion-wake-beam} illustrate the salient features of the non-linear ion-wake. Figure \ref{fig1:ion-wake-laser} shows the excitation phase at an early time when the bubble wake-train is still executing orderly oscillations and its fields have begun to excite inertial ion motion resulting in a soliton-like ion-wake structure ($\delta n_i/n_0 \simeq 0.2$) as seen in Fig.\ref{fig1:ion-wake-laser}(b),(c). At later times as shown in  Fig.\ref{fig2:ion-wake-beam} the radial oscillations sustaining the bubble train have phase-mixed, converting much of the wave energy into electron thermal energy. The resulting electron thermal pressure drives the ion-soliton ($\delta n_i / n_0 > 1$) outwards. The time evolution of the radial dynamics of the ion rings driven by the time-asymmetric nonlinear electron is shown in a movie in supplementary material \cite{supplementary-movie}. We show below, it is the longitudinal or time asymmetry of the radial electron wakefields that excites the ion soliton which propagates leaving behind a near-void channel shown in the PIC simulations in Fig.\ref{fig1:ion-wake-laser} and Fig.\ref{fig2:ion-wake-beam}.

The paper is organized into the following sections. In section \ref{radial-ion-wave} using the linearized fluid equations for ion dynamics we show the two distinct phases of the ion-wake: the excitation phase and the propagation phase. Using the fact that the thermalizing electron wakefield is non-isothermal with radial electron temperature gradients, we model the non-linear ion-acoustic waves as a driven cylindrical ion soliton. We use an analytical model based on fields of a non-linear plasma wave and simulations to demonstrate the inertial phase of the ion-wake in section \ref{ion-soliton-excitation-phase}. In section \ref{ion-soliton-propagation-phase} the propagation phase of the ion-wake is analytical modeled with simulations verifying the propagation of the cylindrical ion-soliton driven by the radial temperature gradient of the phase-mixed electrons. Finally, in section \ref{ion-soliton-positron-wake} we introduce and analyze the properties of ``crunch-in" wakefield regime in an ion-wake channel, using analytical model and simulations. In appendix \ref{ion-wake-model-considerations} we present considerations and assumptions made to derive the ion-wake model.

\section{Non-linear Ion-wake: \\ as a driven Plasma Ion-wave}
\label{radial-ion-wave}

To develop insight into the ion wake physics, we consider the 1-D simplified dispersion relation of the ion-acoustic plane waves, 

\begin{align}
\omega^2 = \frac{c_s^2k^2}{ 1 ~ + ~ (c_s/\omega_{pi})^2 ~ k^2 }
\label{ion-wave-dispersion-relation}
\end{align}

\noindent where, $\omega_{pi}=\omega_{pe}\sqrt{m_e/m_{i}}$ and $c_s = \sqrt{\Upsilon k_BT_{wk}/m_{i}}$ under the collision-less condition, $T^i_{wk}\ll T^e_{wk}$ and $\Upsilon = 1 + 2/f$ is the adiabatic index with $f$ being the degrees of freedom of the ions. 

At early times the ion motion is dominated by inertia, thus ions move over the plasma-ion timescales when driven by time varying and asymmetric fields of non-linear electron plasma-wave. As the ion inertia leads to very small spatial displacement scales $k \rightarrow \infty$, the term $k (c_s/\omega_{pi}) \gg 1$ (where $c_s/\omega_{pi} = \lambda_{De} =\sqrt{\frac{k_BT_e}{4\pi e^2 n_0}}$ is the Debye wavelength). Thus, $\omega \simeq \omega_{pi}$ and the ion-soliton density spikes grow over the plasma-ion frequency timescales, $2\pi \omega_{pi}^{-1}$. The radial electron oscillations sustaining the bubble undergo phase-mixing, the electron trajectories lose orderly motion and thermalize. As the electrons thermalize, the ion motion is driven by thermal pressure of electrons.

When the ions gain significant momentum and start oscillating over larger spatial scales in response to the electron dynamics then $k\lambda_{De} \ll 1$. In this thermally driven phase, the acoustic wave propagation becomes dispersion-less with $\omega = k c_s$.

An ion-acoustic wave growing in amplitude undergoes self-steepening, forming a density spike over much smaller spatial scales; $k$ becomes large while dispersion becomes important. The ion-wake modeled here is non-linear, thus the large $k$ dispersion relation retaining the higher-order terms in $k$ in the Taylor series expansion of eq.\ref{ion-wave-dispersion-relation}, is, $\omega = c_s k ~ - ~ \frac{c_s}{2} ~ \lambda_{De}^2 ~ k^3$. 

Also, at much later times, the ions undergo heating; $T_i$ increases and modifies the sound speed to $c_s = \sqrt{ k_B (\Upsilon_e T_e + \Upsilon_i T_i) / m_{i}}$. 


\subsection{Time-scale separation of Ion-dynamics: \\ a simplified driven linearized ion-fluid model}

\begin{figure}
	\begin{center}
   	\includegraphics[width=5in]{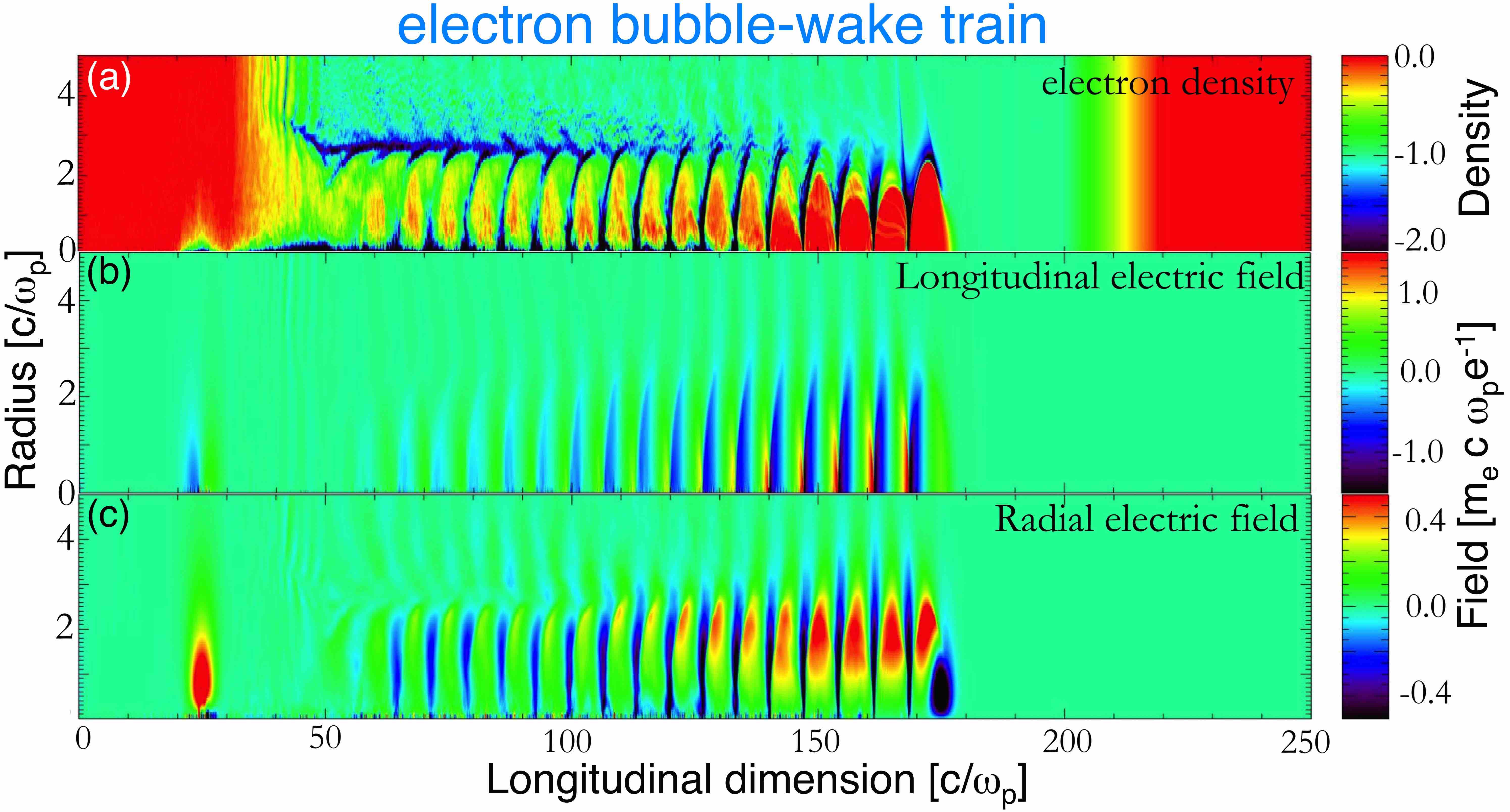}
	\end{center}
\caption{ \small { \bf Bubble-wake train behind an ultra-relativistic electron beam with bubble: $\beta_g \ll \beta_{\phi} \simeq \beta_{beam}$.}   
(a) electron density in 2D cylindrical real-space, (b) corresponding longitudinal electric field profile and (c) corresponding radial-field profile.
Here the beam is located between 170 and 180$\frac{c}{\omega_{pe}}$.
The bubbles just behind the driver in Fig. 3a undergo phase-mixing over several cycles. The intermediate stages of the extent of phase-mixing can be inferred from the bubbles that are closer to the beam. 
The beam-plasma parameters are the same as in Fig.\ref{fig2:ion-wake-beam} but the electron-wake is shown at an earlier time $t = 150\omega_{pe}$. }
\label{fig3:bubble-train}
\end{figure}

In this section, we derive the wave equation for the ion-wake in the linear fluid approximation driven by two terms: the electron wakefields and the electron thermal pressure. The linear ion-acoustic wave can be obtained by perturbative expansion of ion density, $n_i$ and ion fluid velocity $v_i$ in the zeroth-order ion fluid continuity equation, $n_0\mathbf{\nabla}\cdot\mathbf{v}^{(1)}_i + \frac{\partial n^{(1)}_i}{\partial t} = 0$. Taking a partial derivative with time, $\mathbf{\nabla}\cdot\frac{\partial \mathbf{v}^{(1)}_i}{\partial t}  + \frac{\partial^2 }{\partial t^2} \frac{n^{(1)}_i}{n_0} = 0$. The ion-fluid equation of motion where the electron temperature ($T_e$) has a spatial gradient and electron wakefields ($\bf{E}_{wk}$) still persist is, $m_i\frac{\partial \mathbf{v}^{(1)}_i}{\partial t} = eZ_i\mathbf{E}_{wk} - \Upsilon k_BT_e\mathbf{\nabla}\frac{n^{(1)}_i}{n_0} - \Upsilon k_B\frac{n^{(1)}_i}{n_0}\mathbf{\nabla}T_e$. The assumption of spatial gradient of electron temperature has been used because electron plasma wave oscillations phase-mix into non-isothermal plasma (this is substantiated through numerical results in the simulations section in Fig.\ref{fig5:elec-temperature-profile}). Upon substituting the equation of motion in the time-derivative of the linearized continuity equation, $\mathbf{\nabla}\cdot\left(\frac{eZ_i}{m_i}\mathbf{E}_{wk} - \frac{\Upsilon k_BT_e}{m_i}\mathbf{\nabla}\frac{n^{(1)}_i}{n_0} - \frac{ \Upsilon k_B}{m_i} \frac{n^{(1)}_i}{n_0}\mathbf{\nabla}T_e \right)  + \frac{\partial^2 }{\partial t^2} \frac{n^{(1)}_i}{n_0} = 0$. Thus, a driven ion-acoustic wave linearized to the first-order in density perturbation has the form, 

\begin{align}
\nonumber \left(\frac{\partial^2 }{\partial t^2} - c_s^2\nabla^2\right) & \frac{n^{(1)}_i({\mathbf r},t)}{n_0} \\
& = - \frac{eZ_i}{m_i}\mathbf{\nabla}\cdot\mathbf{E}_{wk}({\mathbf r},t) \biggr\rvert_{\mathrm{wake}} + \frac{ \Upsilon k_B}{m_i}\frac{n^{(1)}_i}{n_0}\nabla^2T_e \big\rvert_{\mathrm{thermal}}
\label{ion-wave-equation-first-order}
\end{align}

In this first-order approximate ion-fluid model the right-hand side of eq.\ref{ion-wave-equation-first-order} shows two separate timescales of the ion-wake. 

At earlier times, the first term on the right-hand side dominates. This is the {\it formation or inertial} phase of the ion-wake where the bubble electron oscillations undergo ordered radial motion and the bubble radial electric field excites the inertial response of the ions. The group velocity of the electron bubble wake ($\beta_g\approx 3 v_{th}^2/c^2$, in the 1-D limit, where, $v_{th} \simeq \sqrt{k_BT_e/m_e}$ is the mean electron thermal velocity \cite{Vlasov}) is much smaller than the phase velocity so the bubble fields interact with the background plasma over several oscillations. Fig.\ref{fig3:bubble-train} shows the non-linear electron-wake train (electron density in real space in \ref{fig3:bubble-train}(a)) and its time-asymmetric fields (longitudinal  \ref{fig3:bubble-train}(b) and radial \ref{fig3:bubble-train}(c)) driven by a near speed-of-light energy-source of high-intensity. The fields lead to the formation of the on-axis and $R_B$ ion density spikes. At later times after the phase-mixing between radial oscillators the electrons thermalize and ${\bf E}_{wk}({\bf r},t)\sim 0$. This is the {\it propagation or thermal} phase where the electron thermal pressure gradient drives the cylindrical soliton around $R_B$ radially outwards to many times $R_B$. 

Eq.\ref{ion-wave-equation-first-order} is not directly solved as it can be separated over the two different timescales. In the inertial or excitation phase when the plasma is cold ($T_e\simeq 0$, $c_s\simeq 0$), a better description is provided by an ion-ring model driven by the fields of the electron wave. The ion-ring model is developed and verified using PIC simulations in section \ref{ion-soliton-excitation-phase}. The thermally-driven propagation phase is modeled as a driven ion-acoustic soliton and verified by PIC simulations in section \ref{ion-soliton-propagation-phase}.  

Before proceeding to the solutions in the two time-scales, we illustrate the time-scale separation using PIC simulation snapshot over tens of electron oscillations behind the driver in Fig. \ref{fig3:bubble-train}. In the Fig. \ref{fig3:bubble-train} (a) the coherent motion in the first four or five oscillations behind the driver is evident; whereas further behind the driver, the electrons begin to de-cohere due to phase-mixing. The falling-off of the electron wakefields to nearly zero as the electrons thermalize is later shown using PIC simulations over a much longer time-scale in sec. \ref{ion-soliton-propagation-phase} (see Fig.\ref{fig6:ion-motion-lineouts}(b) where the radial electric field goes to zero around 200$\omega_{pe}^{-1}$ and the ion-soliton is seen moving outwards radially in \ref{fig6:ion-motion-lineouts}(a)).

\section{Excitation phase: \\ Ion Inertial response to the bubble fields}
\label{ion-soliton-excitation-phase}

Since the characteristic time of ion-motion is much longer than the electron oscillations, the longitudinal field ${\mathbf E}_{wk}\cdot\hat{z}$ averages out over the full bubble electron oscillation. So, the ions gain relatively small net longitudinal momentum. However, atypical radial ion-dynamics arise because the radial fields, ${\mathbf E}_{wk}\cdot\hat{r}$ are asymmetric in time as shown in Fig.\ref{fig4:bubble-ion-dynamics} and do not average to zero, driving an average radial ion-momentum. 

\subsection{Ion-ring analytical model: \\ interaction with time-asymmetric bubble radial fields}
\label{excitation-phase-model}

\begin{figure}[ht!]
	\begin{center}
   	\includegraphics[width=5in]{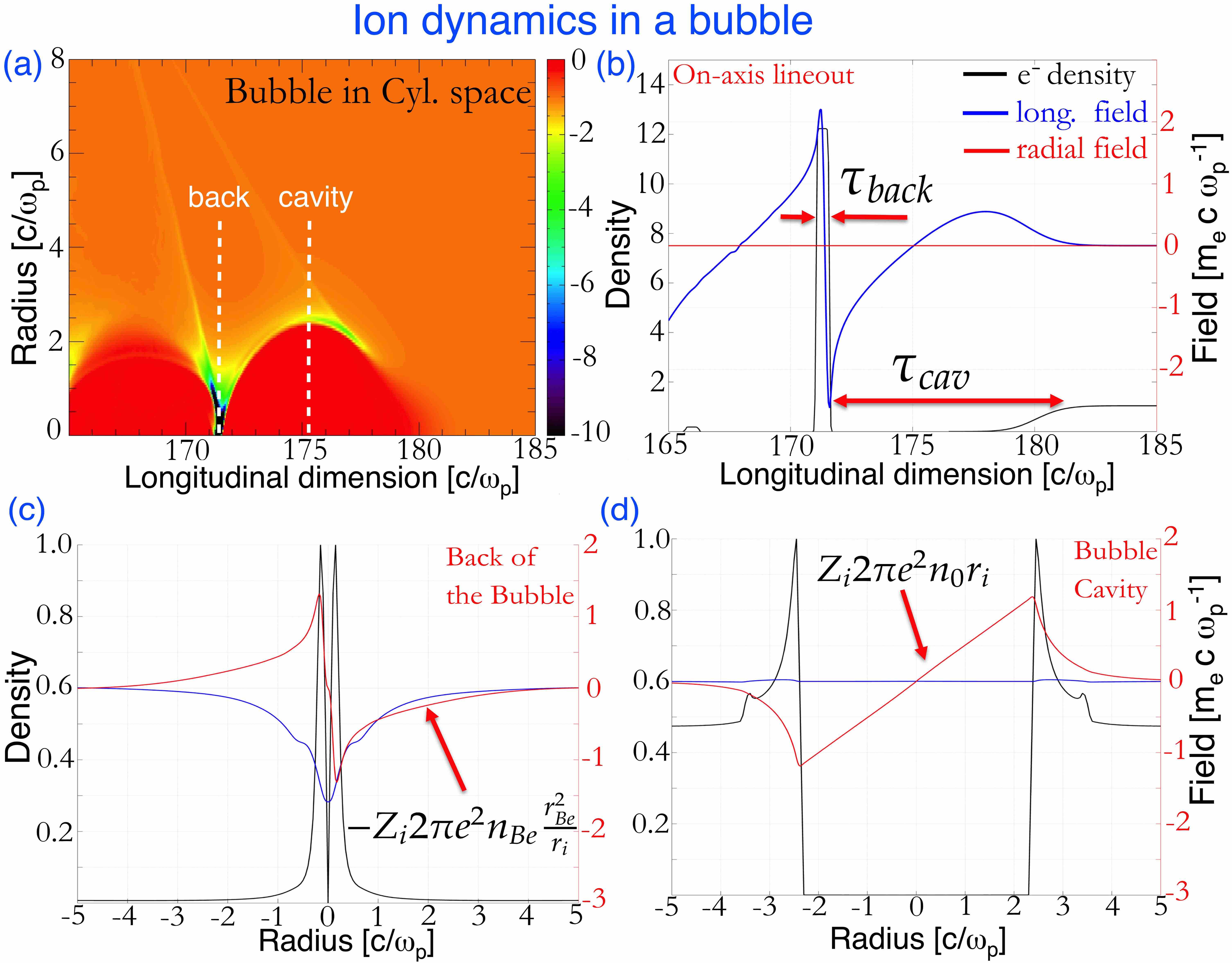}
	\end{center}
\caption{ \small {\bf Ion dynamics in longitudinally asymmetric phases of the radial forces in an electron bubble}. (a) electron density of a bubble in 2D cylindrical real-space. (b) longitudinal on-axis profile of the electron density (black), longitudinal field (blue), focusing field (red). (c) radial-field profile close to the back of the bubble. This is the focussing ``crunch-in" phase for the ions. (d) the fields at the center of the ion-cavity of the bubble. This is the defocussing ``push-out" phase for the ions. } 
\label{fig4:bubble-ion-dynamics}
\end{figure}

The first stage of the ion-wake formation is controlled by the different time-asymmetric phases of ion dynamics inertially responding to the bubble radial field impulses shown in Fig.\ref{fig4:bubble-ion-dynamics} namely, ``suck-in" due to the electron compression in the back of the bubble $F^{back}$ during $\tau_{back}$ shown in Fig.\ref{fig4:bubble-ion-dynamics}(c), and the ``push-out" due to the mutual-ion space-charge Coulomb repulsion force $F^{sc}$ during $\tau_{cav}$ shown in Fig.\ref{fig4:bubble-ion-dynamics}(d). The crunch-in force is spatially-periodic at non-linear plasma wavelength, $\lambda_{Np} \approx 2R_B$ with a duty-cycle $\mathcal{D}=\frac{\tau_{back}}{\tau_{back}+\tau_{cav}}\ll 1$. In addition to the plasma wake, the propagating energy sources themselves impart impulses such as the laser ponderomotive force $F^{pm}\tau_{las}$ ($\tau_{las}$ is laser pulse duration) where $F^{pm}_{e}(r,z) = -\frac{m_ec^2}{2\gamma_e}{\bf \nabla}_{r,z}|{\bf a}(r)|^2$ ($\gamma_e$ is the plasma electron Lorentz factor) and the radial force of the drive beam $F_{b}\tau_{b}$ where $F_{b}(r)=- 2\pi e^2 n_{b} r$. The short driver impulses are neglected (below threshold intensity for direct non-linear ion excitation \cite{ion-motion-intense-beam}\cite{ion-motion-beam-emittance})) because they act on the ions over their sub-wavelength short duration. This is unlike the slowly-propagating wake-plasmon bubbles that undergo continual interaction over many plasma periods. The validity of this assumption is evident from the laser ion-wake in Fig.\ref{fig1:ion-wake-laser}. Since the ponderomotive force of a laser driver is an outward force for both the electrons and ions, the on-axis density-spike cannot be from this direct force from the laser. Similarly the ion-density-spike at the radial wake-edge in an electron beam driven ion-motion cannot be excited directly by the space-charge force of the beam, and is caused by the electron wake's radial-edge density compression.

The Lagrangian fluid model of the ions in a bubble consists of ion-rings under cylindrical symmetry with $m_{i} d^2r_{i}/dt^2 = \Sigma F_{wk}$ (where $F_{wk}$ is the force of the electron wake on the ions). The bared-ion region inside the bubble is assumed to be a positively charged cylinder under steady-state approximation ($R_{B} > r_{Be}$, back of the bubble electron compression radius). The force on the ions from the non-linear electron compression $\delta n_e = n_{Be}\gg n_0$ in the back of the bubble and radius $r_{Be}$, pulls the ion rings towards the axis; and within the bubble, the mutual space-charge force of the ion-rings leads to the ion-rings being driven outwards, away from the axis. The \enquote{suck-in} force on the ions is $F^{back} = - Z_{i} 2\pi e^2 n_{Be} \frac{r^2_{Be}}{r_{i}}$. The space-charge force on the ions in the cavity is $F^{sc} = Z_{i} 2 \pi e^2 n_0 r_{i}$. The equation of motion is $m_{i}d^2r_{i}/dt^2 - \frac{c\beta_{\phi}}{\lambda_{Np}}(F^{sc}\tau_{cav} - F^{back}\tau_{back}) = 0$ using, $\omega_{pi}^2 = Z_{i} ~ 4\pi e^2 n_0 / m_{i}$, we have, 

\begin{align}
\frac{d^2r_{i}}{dt^2} + \beta_{\phi} \frac{\omega^2_{pi}}{2}\left(\frac{n_{Be}}{n_0} \frac{\tau_{back}}{\tau_{cav}} \frac{r^2_{Be}}{r_{i}^2} - 1 \right) r_{i} = 0
\label{ion-ring-motion}
\end{align}

\noindent where we have assumed that $c\tau_{cav}/\lambda_{Np} \simeq 1$. Therefore the ion dynamics is dictated by an equilibrium or a separatrix ion-ring radius, where the inward and the outward impulses balance out, $r_{i}^{eq}=r_{Be}\sqrt{\frac{n_{Be}}{n_0} ~ \mathcal{D} }$. The ion-rings at $r_{i} \le r_{i}^{eq}$ collapse inwards towards the axis resulting in an on-axis density spike. Whereas the ion-rings at $r_{i} \ge r_{i}^{eq}$ move out away from the axis. For $m_{i}/Z_{i}>m_{p}$ the ion-response is slower but similar.

When the radially outward moving ion-rings reach beyond $R_B$, there is excess net negative charge of the wake electrons within the bubble-sheath. As a result the radially propagating ion rings get trapped and start accumulating just inside the bubble-sheath and cannot freely move beyond, forming a density compression at $R_B$. So, the cylindrical ion soliton is formed around $R_B$. This accumulation of the moving ion-rings is shown in Fig.\ref{fig1:ion-wake-laser}, where it is seen that the ion and electron density start forming a peak at $R_B$. 

The radial location of the excitation of the ion-soliton in the non-linear electron-wave regime is much greater than a skin depth, $c/\omega_{pe}$; thus the ion-wake starts off with a spatial-scale which is over several $c/\omega_{pe}$. This is due to the balance of opposing radial forces on the plasma electrons from the driver and the ion cavity, resulting in their radial accumulation at $R_B$ \cite{Pukhov-laser-bubble}. In the laser-driven case - the outward ponderomotive force is balanced by the evacuated ion-cavity: $F^{pm}_{las} = -\frac{m_ec^2}{2\gamma_e}{\bf \nabla}_r \lvert{\bf a}_0(r)\rvert^2 \simeq F_{cav} = 2\pi e^2 n_0 R_{B}$ gives $R_{B} \sim (c/\omega_{pe})^2 \frac{1}{\gamma_e} {\bf \nabla}_r \lvert{\bf a}_0(r)\rvert^2$ when simplified using ${\bf \nabla}_r|{\bf a}_0(r)|^2 \simeq a_0^2/R_{B}$ and $\gamma_e \simeq a_0$, $R_{B} \simeq   \sqrt{a_0} ~ \frac{c}{\omega_{pe}}$ (computationally, $\simeq2\sqrt{a_0} ~ c/\omega_{pe}$ \cite{Lu-bubble-regime}). In the electron beam-driven bubble the outward force of the beam on the plasma electrons is balanced by the inward pull of the evacuated ion-cavity: $F_{b}(R_B) = 2\pi e^2 n_b r_b^2 / R_B \simeq F_{cav} = 2 \pi e^2 n_0 R_B$. This gives, $R_B \simeq \sqrt{\Lambda_b/(\pi n_0)}$, where $\Lambda_b = n_b \pi r_b^2$ is the line charge density of the beam, where $r_b$ is the beam-radius computed here as $2.3~\sigma_r$ to account for $95\%$ of beam particles for a radially Gaussian beam profile.

\subsection{Bubble field time-asymmetry driven ion-soliton: \\ simulation results}
\label{excitation-phase-simulations}

The above ion-ring model is verified using $2\frac{1}{2}D$ OSIRIS PIC simulations \cite{osiris-code-2002} of the ion-wake in the bubble regime by simulating various energy-sources - laser-pulses in cartesian coordinates and electron-beams in cylindrical coordinates.  The laser pulse is circularly polarized with radially Gaussian and longitudinally polynomial profile (as in \cite{laser-pulse-profile}) with $a_0=4$ (not shown $a_0=1.0$ to $40.0$), pulse length of $30\frac{1}{\omega_0}$, matched focal spot-size radius of $40\frac{c}{\omega_0}$, and laser frequency $\omega_0=10\omega_{pe}$ (the pulse dimensions are in the FWHM of the field). The electron beam is initialized with $\gamma_b\sim38,000$, $n_b=5n_0$ (not shown $n_b=0.25n_0$ to $50n_0$) and spatial Gaussian-distribution with $\sigma_r=0.5\frac{c}{\omega_{pe}}$ and $\sigma_z=1.5\frac{c}{\omega_{pe}}$ (the beam spatial dimensions are $5\sigma$ in both the dimensions). The smallest spatial scale, $c/\omega_{pe}$ is resolved in the beam case and $c/\omega_0$ in the laser case (laser frequency $\omega_0$), with 20 cells in the longitudinal direction and 50 cells in the transverse direction. Each of the plasma grid cell has 36 particles. The beam is initialized with 64 particles per cell. The plasma is initialized in the Eulerian specification (non-moving window) and pre-ionized with $Z_i=1$. At the longitudinal boundaries we initialize vacuum space of $50 ~ c/\omega_{pe}$ followed by density ramps of 20 $c/\omega_{pe}$ sandwiching the homogeneous plasma. Absorbing boundary conditions are used for fields and particles.

The electron-beam driven ion-wake soliton structure in theory is compared to the simulations in Fig.\ref{fig4:bubble-ion-dynamics}(a) and \ref{fig3:bubble-train}(a). The observed $R_B=2.45 ~ c/\omega_{pe}$ (just behind the beam) whereas the estimated bubble radius is $R_B = \sqrt{n_b/n_0 ~ (2.3\sigma_r)^2} = 2.57 ~ c/\omega_{pe}$ ($r_b= 2.3\sigma_r=1.15c/\omega_{pe}$, the assumption $r_b\ll R_B$ is not strictly satisfied). In Fig.\ref{fig2:ion-wake-beam} which is in the propagation-phase, the observed ion-wake soliton is located at $r \simeq 3.3 ~ c/\omega_{pe}$ at 460 $\omega_{pe}^{-1}$ which is about $1.7 \frac{2\pi}{\omega_{pi}}$. The ion-soliton is excited at an early time around $R_B$ and in the snapshot in Fig.\ref{fig2:ion-wake-beam} it has propagated outwards. The on-axis density spike drops to a minimum at $r_{i}^{eq} \approx 0.45c/\omega_{pe}$ in Fig.\ref{fig2:ion-wake-beam} whereas the estimated $r_{i}^{eq} = 0.5 ~ c/\omega_{pe}$ ($n_{Be}/n_0 \simeq 12$, $\mathcal{D} \simeq 0.1$, $r_{Be}\simeq0.5 ~ c/\omega_{pe}$). The radial ion momentum $p_r-r$ phase-space in Fig.\ref{fig7:radial-phase-spaces}(b) shows the ions accumulate at the axis and the channel edge, at a time corresponding to Fig.\ref{fig2:ion-wake-beam}(b). The ions at the channel edge are seen to have a drift velocity and a thermal spread. The radial electron momentum $p_r-r$ phase-space in Fig.\ref{fig7:radial-phase-spaces}(a) shows that a large density of thermalized electrons are trapped within the ion soliton which is confirmed from the density plots in Fig.\ref{fig2:ion-wake-beam}(a).

In the laser-driven bubble simulations the expected and observed $R_B \simeq 4 ~ c/\omega_{pe}$ as shown in Fig.\ref{fig1:ion-wake-laser}(a). In Fig.\ref{fig1:ion-wake-laser}(c) the ion-wake soliton is created at  $r=4.2 ~ c/\omega_{pe}$. The expected and observed on-axis density-spike radius is $r_{i}^{eq} = 0.45 ~ c/\omega_{pe}$ ($n_{Be}/n_0 \simeq 8$, $\mathcal{D} \simeq 0.1$, $r_{Be}\simeq0.5 ~ c/\omega_{pe}$). The model for the excitation of this structure of the non-linear wake has been verified for a range of laser and beam parameters from quasi-linear to strongly non-linear electron wake regime.

\section{Propagation phase: \\ Soliton driven by electron thermal pressure gradient}
\label{ion-soliton-propagation-phase}

As described in section \ref{ion-soliton-excitation-phase} the electron bubble-wake train fields excite a cylindrical ion soliton. Eventually, the electron oscillations phase-mix and thermalize as electron thermal energy on the time-scale of about an ion plasma period. In this section we model the propagation of the cylindrical soliton radially outwards driven by the temperature gradient as shown in eq.\ref{Cylindrical-KdV-equation}. This soliton propagation is modeled using a modified cKdV equation in a non-equilibrium condition such that an electron temperature gradient sustains and drives the cylindrical ion soliton.

\subsection{Thermally-driven ion-acoustic soliton: \\ Analytical model}
\label{driven-soliton-model}

In the linear regime the homogenous ion-acoustic wave equation predicts sinusoidal radial ion oscillations that support the wave. However, the linearized ion-acoustic wave equation is inadequate to describe the propagating solitary density spike at the ion-wake edge, with ion density accumulation many times the background density.  


When the density in the ion perturbation begins to rise to the order of the background density, the electrostatic potential due to charge-separation between the ions and the thermal electrons correspondingly rises. This leads to {\it wave-steepening} due to the preferential acceleration of ions in the direction of the ion-acoustic wave velocity. When the potential of the wave is large enough the ions get trapped and co-propagate at the ion-acoustic wave phase velocity, this non-linearity is the basis of the soliton. It should be noted that the linearized kinetic theory does not formally incorporate the trapping of particles at the wave phase-velocity. In this limit the density perturbation shape is therefore not sinusoidal as the co-propagating background ions accumulate and their density perturbation takes the form of an ion-soliton. The co-propagating ion velocity in the soliton can therefore exceed the ion-acoustic phase velocity, $v_i > c_s$ and $\mathcal{M}-1 > 0$ where $\mathcal{M} = v_i/c_s$ is the Mach number. Therefore, non-linear acoustic waves are in the form of a soliton and propagate faster than the ion-acoustic velocity. 

To second-order, the non-linear ion-density spike $n_{i}(r,t) > n_0$ propagation is governed by the Korteweg-de Vries (KdV) equation \cite{KdV-non-linear-ion-waves} which has propagating solutions of the form $\mathcal{U}(r-\mathcal{M}c_st)$ \cite{Berezin-Karpman-1964} where $\mathcal{U}$ is the ion-acoustic waveform, a soliton solution and $\mathcal{M}$ is the Mach number ($= v_i/c_s$) of the propagating solution. Higher-order contributions to the KdV equation have also been considered by earlier works. However, the more important and relevant here is that the standard form of the cKdV equation assumes an isothermal plasma whereas the bubble-wake phase-mixes into a plasma with a radial electron temperature gradient, whereas the ions are initially cold. In a non-isothermal plasma the effect of trapped electrons in the ion-soliton have been considered using the Bernstein-Greene-Kruskal (BGK) model at the ion-acoustic velocity \cite{Schamel-trapped-particle-modes}.

It is also known that a single ion-soliton under the appropriate conditions can break-up into multiple solitons leading to a N-soliton solution. This is also a phenomenon we observe in the simulations shown in the ion density of the beam-driven case at $z = 60\frac{c}{ \omega_{pe} }$ in Fig.\ref{fig2:ion-wake-beam}.

\begin{figure}[ht!]
	\begin{center}
   	\includegraphics[width=5in]{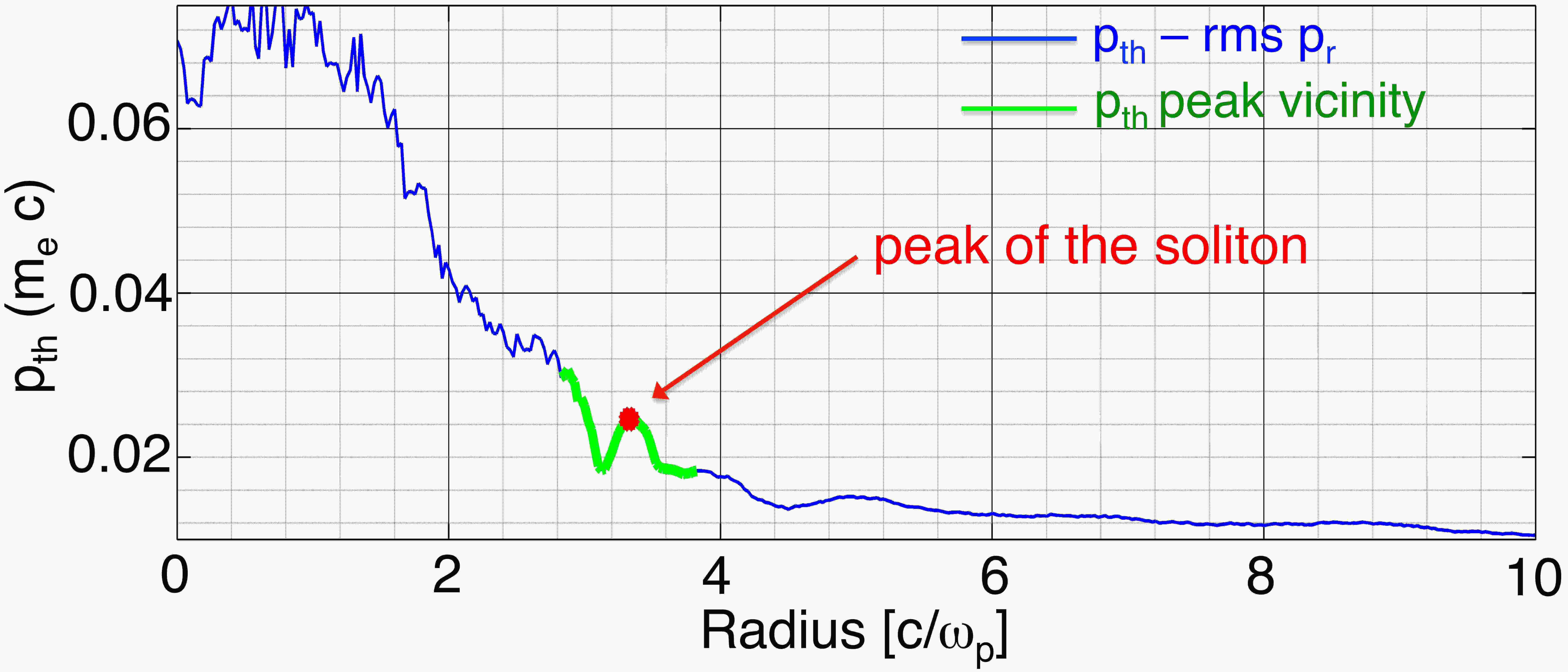}
	\end{center}
\caption{ \small {\bf Radial profile of the root-mean-square radial electron momentum (proportional to the square-root of the electron temperature, $\sqrt{T_e}$) at 460 $\omega_{pe}^{-1}$ for the beam-driven ion-wake in Fig.\ref{fig2:ion-wake-beam}}. The blue curve shows the root-mean-squared radial electron momentum, $p^e_{th}(r) = \sqrt{ \left[ \Sigma_k ~ p^2_r(k,r) ~ 2\pi r ~ n_e(k,r) \right] / \Sigma_{k} ~ 2\pi r ~ n_e(k,r) }$, profile of the wake electrons corresponding to the time in Fig.\ref{fig2:ion-wake-beam} at 460 $\omega_{pe}^{-1}$. This represents the square-root of the electron temperature, $p^e_{th} ~ \propto ~ \sqrt{T_e}$. The radial gradient of the temperature, $\frac{ \partial }{ \partial r } T_e$ is thus computed at the peak of the soliton (red) and in its vicinity (green). It is interesting to note that $\frac{\partial }{\partial r}T_e^{(1)}\biggr\rvert_{peak} = 0$. }
\label{fig5:elec-temperature-profile}
\end{figure}

We consider a description of the non-linear cylindrical ion-acoustic waves with a radial temperature gradient. We assume that the background electron trapping does not significantly modify the distribution function. We assume that the temperature changes slowly in vicinity of the ion soliton. This assumption is validated by the PIC simulations in Fig.\ref{fig5:elec-temperature-profile}.

To obtain the KdV equation \cite{Sahai-dissertation} in cylindrical coordinates with radial temperature gradient we normalize with respect to the local electron temperature, the radius: $\hat{r} = \frac{r}{\lambda_D}$, time: $\hat{t} = \omega_{pi}t = t\sqrt{\frac{4\pi e^2 n_0}{m_i}}$, electric field: $\hat{E} = \frac{e \lambda_{De} }{k_B T_e} E$, potential: $\phi = \frac{e}{k_B T_e} \Phi$, ion-density perturbation: $\hat{n}_i = n_i/n_0$, electron-density perturbation: $\hat{n}_e = n_e/n_0$, ion-fluid velocity: $\hat{v}=\frac{v_i}{c_s}$. Under this normalization the cylindrical coordinate equations transform as: electron Boltzmann distribution equation $\frac{\partial \hat{n}_e}{\partial \hat{r} } = -\hat{n_e} \hat{E} - \hat{n}_e \phi \frac{ \partial }{ \partial \hat{r} } ~ \mathrm{ln} T_e$, ion-fluid continuity equation $\frac{\partial }{\partial \hat{t} } \hat{n}_i + 2 \frac{ \hat{n}_i \hat{v}}{\hat{r}} + 2\frac{\partial}{\partial \hat{r}} \hat{n}_i \hat{v} = 0$, ion-fluid equation of motion $\frac{\partial }{\partial \hat{t} } \hat{v} + \hat{v} \frac{\partial}{\partial \hat{r}} \hat{v} = \hat{E}$ and the Poisson equation $\nabla^2\Phi = \frac{1}{\hat{r}} \frac{\partial}{\partial \hat{r}} (\hat{r}\hat{E})  = \hat{n}_i - \hat{n}_e$. The electric field $\hat{E}$ is both due to the thermal pressure and the radial fields of the wake, $\hat{E}_{wk} + \hat{E}_{th}$. But, in the following analysis the propagation of a non-linear ion-acoustic wave is considered, so we assume that the electron oscillations are thermalized and thus the effect of the fields of the wake is negligible, $\hat{E}_{wk} \rightarrow 0$.

We look for a propagating disturbance of $\hat{n}_e$, $\hat{n}_i$, $\hat{v}$ and $\hat{E}$ in a stationary background plasma with uniform background density $n_0$. We consider weakly non-linear ion-acoustic wave and expand all the wave quantities in the powers of $\delta = \mathcal{M} - 1$. We perturbatively expand $\hat{n}_i$, $\hat{n}_e$, $\hat{E}$, $\phi$, $T_e$ and $\hat{v}_i$ and retain all terms up to the order of $\delta^2$. Note that we have assumed that before the electron wake excitation the plasma is cold, $T_e^{(0)} \simeq 0$.

We transform to a moving frame of the steepened ion density perturbation using the coordinate transform $\xi=\delta^{1/2}(\hat{r} - \hat{t})$ and $\tau = \delta^{3/2}\hat{t}$. Using this, $\hat{r} = \delta^{-1/2}(\xi + \delta^{-1}\tau)$ and $\frac{\partial \xi}{\partial \tau} = \frac{\partial \xi}{\partial \hat{t} } \frac{\partial \hat{t} }{\partial \tau} = -\frac{1}{\delta}$. We renormalize the electric field as, $\tilde{E} = \delta^{-1/2}\hat{E}$. Note that in the moving frame the potential gradient is, $E = -\frac{\partial }{\partial \hat{r}} \Phi = -\delta^{1/2} \frac{\partial }{\partial \xi} \Phi $, so $\tilde{E}$ is a more appropriate quantity.

Under the assumption that in the moving-frame the quantities of the disturbance change with small $\delta = \mathcal{M} - 1$, the terms in equations are perturbatively expanded and the terms with same powers of $\delta$ are collected. From the $\delta^{1}$ order terms of all the equations above, we infer $\Phi^{(1)} = n^{(1)}_{e} = v^{(1)} = n^{(1)}_{i}\equiv \mathcal{U}$ and $\frac{\partial }{\partial \xi}\mathcal{U} = -\tilde{E}^{(1)}$. 

By collecting the $\delta^{2}$ terms from the Boltzmann's equation we obtain $\tilde{E}^{(2)} = -\frac{\partial }{\partial \xi} n^{(2)}_{e} + \mathcal{U}\frac{\partial }{\partial \xi}\mathcal{U} - \mathcal{U}\frac{\partial }{\partial \xi}T_e^{(1)}$. Similarly, collecting the $\delta^{2}$ terms from the ion-fluid equation of motion we obtain $\frac{\partial }{\partial \xi} \hat{v}^{(2)} - \frac{\partial }{\partial \xi} n^{(2)}_{e} = \frac{\partial }{\partial \tau} \mathcal{U} + \mathcal{U} \frac{\partial }{\partial \xi} T_e^{(1)}$ and from the Poisson equation we obtain $\frac{\partial^3 }{\partial \xi^3} \mathcal{U} = - \frac{\partial }{\partial \xi}(n^{(2)}_{i} - n^{(2)}_{e})$. Taking the $\delta$-order terms of the continuity equation and substituting $\mathcal{U}$ we obtain, $\mathcal{U} + \tau \left( \frac{\partial }{\partial \tau} \mathcal{U} + 2\mathcal{U}\frac{\partial }{\partial \xi}\mathcal{U} + \left[ \frac{\partial }{\partial \xi} v^{(2)} - \frac{\partial }{\partial \xi} n^{(2)}_{e} \right] \right) - \delta \left(U^2 + v^{(2)} \right) = 0$.  Neglecting quantities with $\delta$ times the second-order terms and using the $\frac{\partial }{\partial \tau} \mathcal{U}$ result above, $\frac{\mathcal{U}}{\tau} + 2\frac{\partial }{\partial \tau} \mathcal{U} + 2\mathcal{U}\frac{\partial }{\partial \xi}\mathcal{U} + \left[\frac{\partial }{\partial \xi} n^{(2)}_{e} - \frac{\partial }{\partial \xi} n^{(2)}_{i}\right] = - \mathcal{U}\frac{\partial }{\partial \xi}T_e^{(1)}$. 

Using the $\delta^2$ terms of the Poisson equation in the above result and using the self-similarity property of the ion-soliton, we obtain the driven Korteweg-de Vries equation in cylindrical coordinates \cite{Maxon-cyl-soliton} (a more detailed derivation of this modified cKdV model can be found in \cite{Sahai-dissertation}), 

\begin{align}
%
\nonumber & \Phi^{(1)} = n^{(1)}_{e} = v^{(1)} = n^{(1)}_{i}\equiv \mathcal{U} \\
& \frac{\mathcal{U}}{\tau} + 2\frac{\partial }{\partial \tau} \mathcal{U} + 2\mathcal{U}\frac{\partial }{\partial \xi}\mathcal{U} + \frac{\partial^3 }{\partial \xi^3} \mathcal{U} = - \mathcal{U}\frac{\partial }{\partial \xi}T_e^{(1)}
\label{Cylindrical-KdV-equation}
\end{align}

It differs from the cartesian-KdV equation by the term $\frac{\mathcal{U}}{\tau}$ and the temperature-gradient based driver term $- \mathcal{U}\frac{\partial }{\partial \xi}T_e^{(1)}$. The cartesian KdV equation can be analytically solved to obtain two classes of solutions: (a) self-similar solutions which are shown in \cite{Berezin-Karpman-1964} to be Airy functions and (b) soliton solutions. A \enquote{soliton} is a single isolated pulse which retains its shape as it propagates at some velocity, $v_{soliton}$. This means that for a soliton-like solution the $\mathcal{U}$ only depends upon the soliton-frame variable, $\zeta = \xi - \mathcal{M}c_s \tau$ and not on space-like $\xi$ and time-like $\tau$ variables separately. The solution of the cartesian KdV equation in this co-moving frame is $\mathcal{U}(\zeta)=3 v_s ~ \mathrm{sech}^2(\sqrt{ \frac{v_s}{2} } \zeta)$ \cite{Berezin-Karpman-1964}.

The cKdV equation and the driven cKdV equation obtained here cannot be solved analytically. However, the numerical analysis and experimental verification \cite{cyl-soliton-observation} of the cylindrical-KdV (cKdV) equation show that it supports functions of the form $\mathcal{U} \propto \mathrm{sech}^2(r-\mathcal{M}c_st)$ in the form of a cylindrical ion-soliton. But, the amplitude of the cylindrical soliton changes as it propagates. Here, we show that the wake electron temperature drives the ion soliton for much longer distances than possible in an isothermal plasma. The velocity of the soliton in the cylindrical case is higher than in cartesian case \cite{Maxon-cyl-soliton}. Since the ion-wake is excited in a non isothermal plasma its velocity changes as it is driven. The mean electron temperature reduces as the soliton propagates radially outwards because the electron thermal energy is distributed over a larger volume. The cKdV equation is also known to support an N-soliton solution, and simulations show N-soliton forming during the propagation phase. We computationally seek the dependence of the non-linear ion-density spike on $(r-\mathcal{M}c_st)$-coordinate. 

It should be noted that such soliton solutions are supported under certain limiting condition on the Mach number, $\mathcal{M}$. The strict condition on the existence and stability of ion-soliton arises from a threshold limit on the magnitude of soliton potential to continue trapping the background ions. 

Here we find that the speed of ion soliton is nearly equal to and only slightly higher that the ion-acoustic speed calculated using the mean temperature. As this is not an isothermal plasma, there is no well-defined ion-acoustic speed. So, the ion-acoustic wave is phase-mixed and its velocity also changes as it propagates. 

The local electron temperature of the ion soliton as shown in Fig.\ref{fig5:elec-temperature-profile}, is used to calculate the Mach number, $\mathcal{M}$ and thus a stability criterion can be derived. This problem is represented using the condition on the Sagdeev psuedo-potential, $\mathcal{V}(\phi) = - \left( \mathrm{exp}(\phi) -1 + \mathcal{M}(\mathcal{M}^2 - 2\phi)^{1/2} - \mathcal{M}^2 \right)$ that it has to be a real number. This condition is satisfied when $\mathcal{M}^2 - 2\phi \geq 0$ therefore, $\phi < \mathcal{M}^2/2$ and $\phi_{max} = \mathcal{M}^2/2$. Using this we find the well-known condition, 

\begin{align}
%
\nonumber & 1 < \mathcal{M} < 1.6, \quad v_i < 1.6 ~ c_s \\
& \Phi < \frac{\mathcal{M}_{max}^2}{2} = 1.28 
\label{KdV-solution-existence}
\end{align}

As will be shown later, we find from simulations that the Mach number calculated using the mean temperature is well within these bounds, and thus the soliton is stable.


\subsection{Thermally-driven ion-acoustic soliton: \\ simulation results }
\label{driven-soliton-simulations}

\begin{figure}[ht!]
	\begin{center}
   	\includegraphics[width=5in]{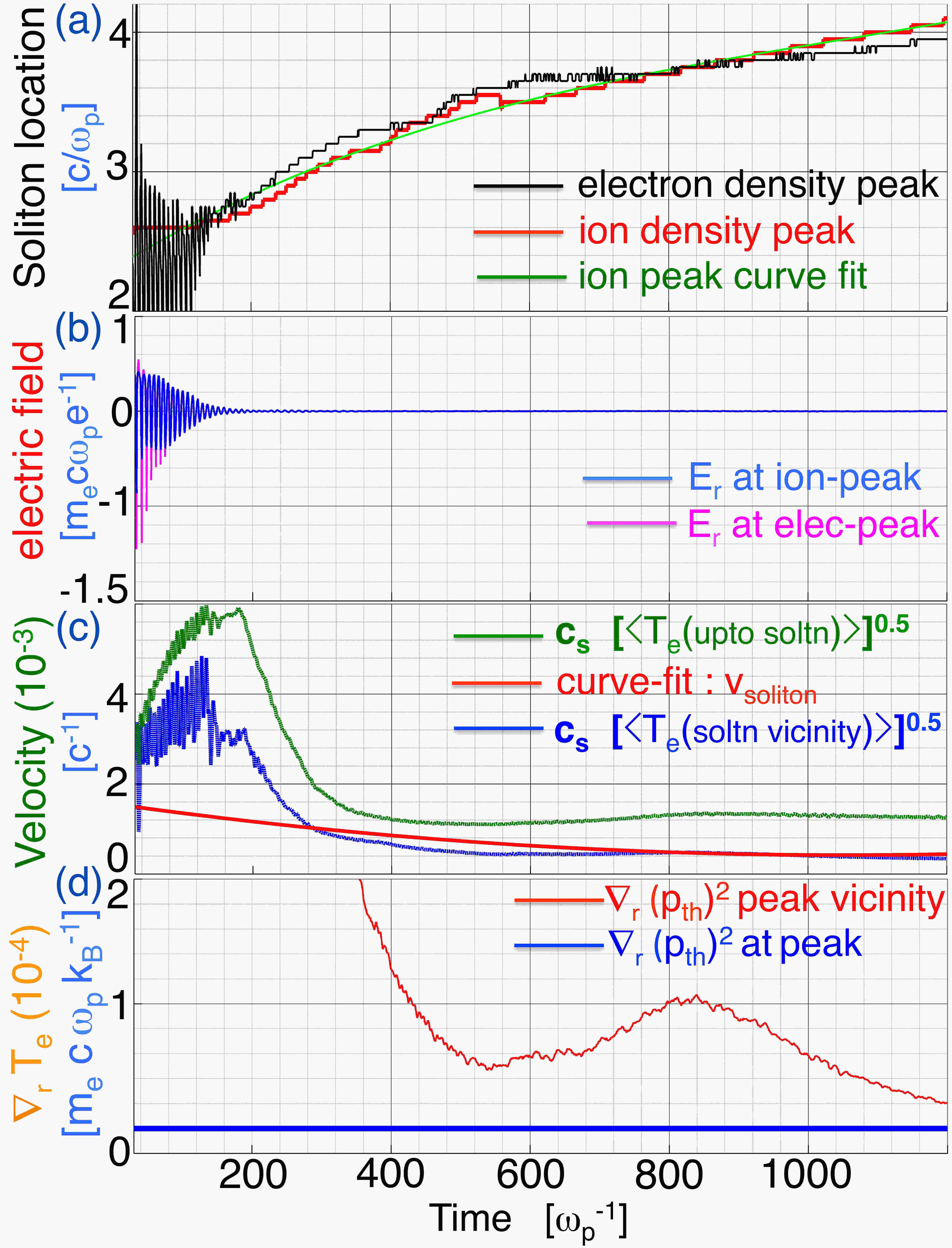}
	\end{center}
\caption{ \small {\bf Time evolution of the cylindrical ion soliton}. (a) electron (black) and ion (red) spike radial positions (in terms of $c\omega_{pe}^{-1}$) with time and a third-order fit (green) for the position of the ion density-spike of the soliton. (b) radial wakefields of the electron bubble oscillations (in terms of $m_e c \omega_{pe}e^{-1}$) at the electron density spike (magenta) and at the ion density spike (blue). (c) radial velocity of the ion density spike of the soliton calculated from the third-order fit curve (red). An estimate of the sound speed (green) using the mean temperature, between the axis \& the soliton location (green) and in the vicinity of the soliton peak (blue), in the expression $c_s = \sqrt{k_B \langle T_e \rangle / m_i}$. Since the plasma is not isothermal the mean temperature is calculated by averaging the temperature of electrons over the indicated spatial region. (d) gradient of the electron temperature at the soliton ion density peak (blue) and in the vicinity of the peak (red). The vicinity of the ion density peak of the soliton is defined as shown in Fig.\ref{fig5:elec-temperature-profile}. }
\label{fig6:ion-motion-lineouts}
\end{figure} 

The channel-edge density spike, with a form similar to the cKdV-solution in the $r-\mathcal{M}c_st$ frame as shown in Fig.\ref{fig1:ion-wake-laser} and Fig.\ref{fig2:ion-wake-beam} is seen to be propagating radially outwards. The propagation phase starts around $t = 200 \omega_{pe}^{-1}$ as the radial electric fields $\mathbf{E}_{wk} \rightarrow 0$ as shown in Fig.\ref{fig6:ion-motion-lineouts}(b). The propagation phase is evident in Fig.\ref{fig6:ion-motion-lineouts}(a) where the red curve is the position of the peak of the ion-soliton in time. The radial position of the peak of the ion-soliton from each of the PIC electrons and ion density snapshot is obtained in the post-processing scripts and this is shown in Fig.\ref{fig6:ion-motion-lineouts}(a). The cylindrical ion-soliton has propagated from  $r_{soliton}(460 \omega_{pe}^{-1}) = 3.3 c/\omega_{pe}$  (also seen in Fig.\ref{fig2:ion-wake-beam}) to $r_{soliton}(1100 \omega_{pe}^{-1}) = 4.1 c/\omega_{pe}$ which corresponds to an average speed of $\langle v_{soliton} \rangle = 0.0013c$. 

We compare the time-averaged soliton speed $\langle v_{soliton} \rangle$ to the average speed of sound, $c_s/c = p^{e}_{th} \sqrt{\frac{\Upsilon}{2}\frac{m_e}{m_i}}$ where the average $p^{e}_{th} \simeq 0.06$ from the electron phase-space (not shown). This gives $c_s \simeq 0.001c$ ($\Upsilon = 2$ for 2D) in agreement with the average soliton velocity. Using this time-averaged analysis we see that $\mathcal{M} \simeq 1.3$ and so the stability criteria in eq.\ref{KdV-solution-existence} is satisfied. 

However, as the soliton moves out the volume between the axis and the soliton edge increases. Thus the electron thermal energy re-distributes and spreads over a larger volume. This leads to the reduction in the temperature with time. The soliton is not freely propagating but is driven by the radial gradient of the electron temperature as shown in eq.\ref{Cylindrical-KdV-equation}. The soliton speed thus changes in time as shown in the red curve of Fig.\ref{fig6:ion-motion-lineouts}(c). The sound speed also varies with time and it is estimated using the temperature at that instant using $c_s(t) = \sqrt{k_B \langle T_e(t) \rangle / m_i}$. The plasma is not in thermal equilibrium and its temperature varies radially as shown in Fig.\ref{fig5:elec-temperature-profile}. The root-mean-square radial momentum is used to estimate the temperature at an instant of time, and is calculated over radial dimension from the $p_r-r$ phase-space. The mean temperature is calculated by taking the average of the rms radial momentum - over the entire channel: channel-$\langle p^e_{th} \rangle = \left[ \Sigma^{r_{sol}}_{r=0} ~ p_{th}(r) ~ 2\pi r ~ n_e(r) \right] / \Sigma^{r_{sol}}_{r=0} ~ 2\pi r ~ n_e(r) ~ \propto ~ \mathrm{channel} - \sqrt{\langle T_e \rangle }$ or in the vicinity of the soliton: soliton-$\langle p^e_{th} \rangle = \left[ \Sigma^{r_{sol}+\epsilon}_{r_{sol}-\epsilon} ~ p_{th}(r) ~ 2\pi r ~ n_e(r) \right] / \Sigma^{r_{sol}+\epsilon}_{r_{sol}-\epsilon} ~ 2\pi r ~ n_e(r) ~ \propto ~ \mathrm{soliton} - \sqrt{\langle T_e \rangle }$. The instantaneous sound speed, $c_s(t)$ computed with channel-$\langle p^e_{th} \rangle$ is in the green curve in Fig.\ref{fig6:ion-motion-lineouts}(c) and $c_s(t)$ computed with soliton-$\langle p^e_{th} \rangle$ is in the blue curve in Fig.\ref{fig6:ion-motion-lineouts}(c). The extent of the vicinity ($\epsilon$) around the soliton peak is shown in Fig.\ref{fig5:elec-temperature-profile}.

We compare the curves in (i) red: $v_{soliton}(t)$ (from the 3rd-order polynomial curve-fit of the radial position of the ion-density peak as a function of time), (ii) green: $c_s(t)$ from channel-$\langle p^e_{th}(t) \rangle$ and (iii) blue: $c_s(t)$ from soliton-$\langle p^e_{th}(t) \rangle$ in Fig.\ref{fig6:ion-motion-lineouts}(c). From the comparison it is observed that they are in good agreement. It can be seen that the velocity of the soliton estimated using the location of the ion-density peak (red) lies between $c_s(t)$ calculated using the average temperature over the channel (green) which is the upper limit and $c_s(t)$ calculated using the average temperature over the soliton (blue) which is the lower limit. 

We also present the radial gradient of the electron temperature, $\frac{ \partial }{ \partial r } T_e(r,t)$ in Fig.\ref{fig6:ion-motion-lineouts}(d). It is interesting to note from the blue curve in Fig.\ref{fig6:ion-motion-lineouts}(d) that the temperature gradient at the peak of the ion-soliton is zero, $\frac{ \partial }{ \partial r } T_e(r,t) \big\rvert_{peak} = 0$. In the vicinity of the soliton peak we see that the gradient of the temperature follows the variation in the ion soliton velocity, this follows from eq.\ref{Cylindrical-KdV-equation}. The vicinity of the soliton peak is shown as the green curve overlaid on the thermal momentum curve in Fig.\ref{fig5:elec-temperature-profile}.

In Fig.\ref{fig2:ion-wake-beam}(b) N-soliton formation is observed in the ion-density at around $z \simeq 60 \frac{c}{\omega_{pe}}$. The single-ion soliton is seen splitting into several solitons. The N-soliton solution can explained by the seeding of different initial momentum of the ion-rings because ion-rings driven in the ``push-out" phase have a radial position dependent defocussing force acting on them, $F^{sc}(r_{i}) = Z_{i} 2 \pi e^2 n_0 r_{i}$. This is shown in Fig.\ref{fig4:bubble-ion-dynamics}(d). Thus the ion-rings originating at a larger radii from the axis are pushed outwards with a force of a higher magnitude and the rings originating at a smaller radii just outside $r_i^{eq}$ are pushed outwards by a smaller force. So, over a longer time the set of ion-rings with a higher initial momentum propagate radially outwards at a larger radial velocity. This break-up of a single ion-soliton into N-solitons occurs over a longer time-scale because the difference in momentum is small compared to the average momentum. 

The thermal momentum, $p_{e}^{th}$ at this time is less than one-tenth of the peak wake quiver momentum. There are several reasons for the cooling, such as, transfer of the wake energy to the ions and the trapped electrons \cite{phase-mixing-trapping}, escape of the highest energy electrons and un-trapped ions from the channel edge, energy loss to the bow-shock and the re-distribution of the energy over an expanding volume. The peak radial ion-momentum is $\simeq0.005$ which shows that not all the radially propagating ions are trapped. The un-trapped free-streaming ions at $\simeq 7c/\omega_{pe}$ can be distinguished from the ions at the channel-edge in $p_r-r$ phase-space.

\begin{figure}[ht!]
	\begin{center}
   	\includegraphics[width=5in]{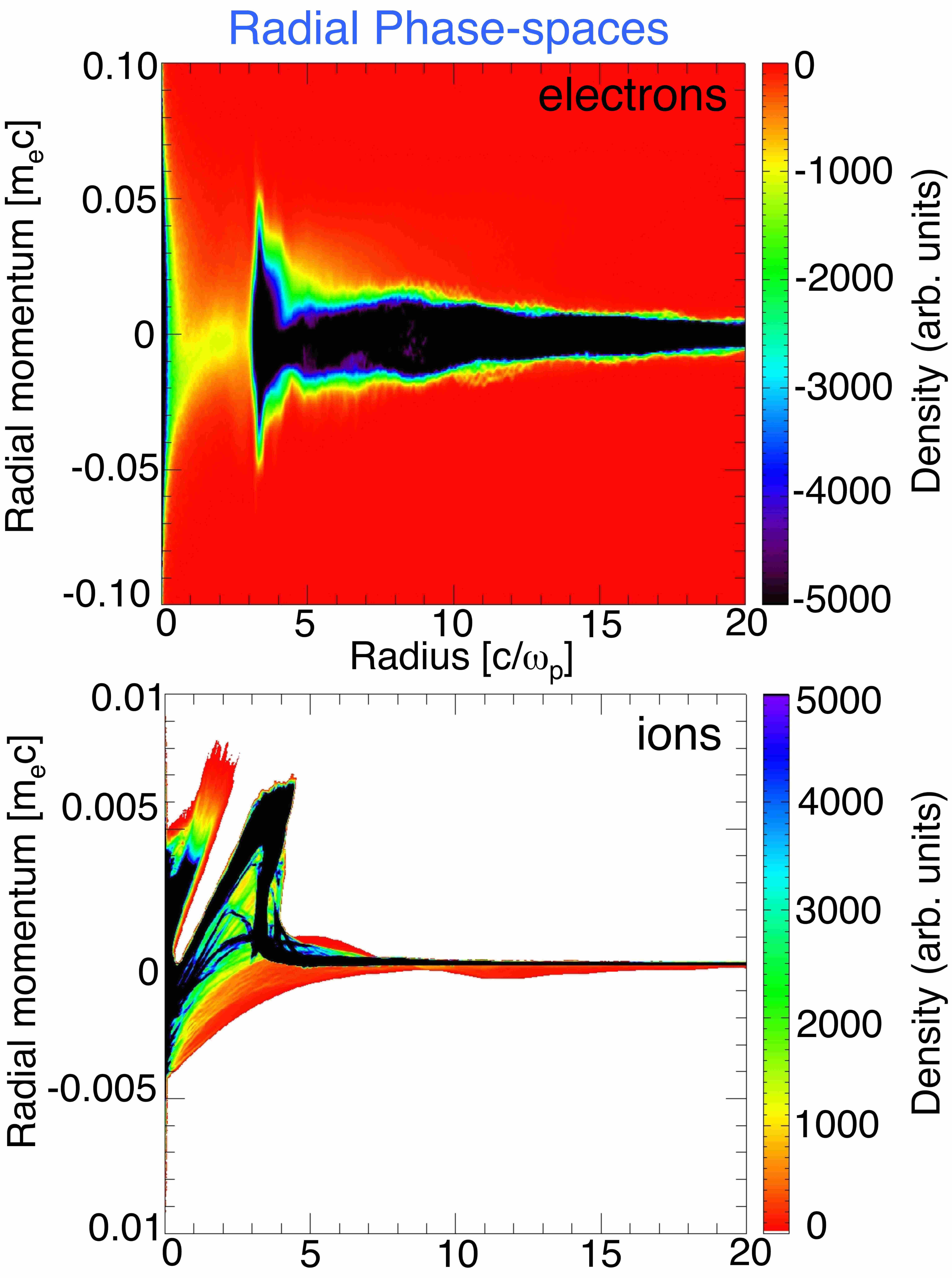}
	\end{center}
\caption{ \small {\bf Radial phase-space snapshots of the electron and ion density in Fig.\ref{fig2:ion-wake-beam}}. (a) electron $p_r-r$ radial momentum phase-space showing the accumulation of thermalized electrons within the ion-soliton. (b) ion $p_r-r$ radial momentum phase-space showing the on-axis and ion-wake edge ion accumulations. }
\label{fig7:radial-phase-spaces}
\end{figure}

It should be noted that the long-term stability of the on-axis ion-density spike of the non-linear ion-acoustic wave is not fully modeled here. The on-axis ion-density spike will disintegrate due to mutual Coulomb repulsion of the ions over the sub skin-depth spike radius. This effect of the collapse of the on-axis density-spike will be addressed in future work. We expect that the disintegration of the central structure to be further by azimuthal asymmetries not included in the cylindrically symmetric simulations. Earlier disintegration is seen in cartesian simulations not shown.

In summary, the ion-wake is a near-void channel with sub-skin-depth density-spikes on-axis and at the bubble-edge located at the bubble-radius, $R_{B}$ \cite{Pukhov-laser-bubble} of several $c/\omega_{pe}$. The ion accumulation in both the density-spikes is many times the background density, and the outre spike propagates outwards as a solitary structure at slightly above the speed of sound. 

The time-scale of dissipation of ion-wake and relaxation of the plasma distribution to $v_{th}/c\sim 0$ sets an upper limit on the repetition-rate \cite{hot-plasma-wake} of the future plasma colliders. It is well known that the ion acoustic wave is damped by collisions and ion-wave Landau damping. The ion dynamics opens  questions upon the plasma container walls and the distance needed to avoid damaging them by the significant radially outward ion-flux. 

It was earlier suggested that the wakefield energy in the plasma wave could be replenished and sustained \cite{beam-wakefield-SLAC} by a train of energy source in order to achieve high repetition rate. However, as shown in this paper due to ion motion this is not possible.

\section{Positron acceleration: \\ ``Crunch-in" regime in the Ion-wake channel}
\label{ion-soliton-positron-wake}

We explore the use of the ion-wake channel for positron-beam driven positron wakefield acceleration in a novel and relevant \enquote{crunch-in} regime where the channel radius is of a few $c/\omega_{pe}$ as is the case for the ion-wake channel. Such channels are also promising \cite{hollow-channel-chiou}\cite{hollow-acclerator-1998}\cite{hollow-accelerator-2013} for exciting the well-studied purely electromagnetic electron-wakefields. These pure electromagnetic fields driven in a hollow-channel have proven to have zero focusing forces when driven by relativistic particles \cite{hollow-acclerator-1998}. Here we show that in the Crunch-in regime driven even in an ion-wake channel, strong accelerating and focusing fields of electrostatic nature are excited by the electron rings crunching in from the channel wall.

The ion-wake enables the \enquote{Crunch-in} regime because as it slowly propagates radially outwards the channel radii scans over a variety of $c/\omega_{pe}$, while its length is the energy-source plasma interaction length. Meter-scale propagation of electron beams and few centimeter-scale propagation of laser beams in plasmas while exciting nonlinear electron-waves has been well characterized in experiments. The theoretical model presented above thus provides a mechanism to generate long channels of several skin-depth radii. As we show below, the non-linear \enquote{crunch-in} regime requires such channels to optimally match with the driving energy-source.

It is well known \cite{positron-accln-2001} that in a homogeneous plasma positron beam driven wakes have two major problems \cite{positron-IPAC-2015} - (i) The plasma electrons collapsing to the axis from different radii arrive at different times, preventing optimal compression. This is because the radial force of the positron beam driving the ``crunch-in" decreases with the radii. (ii) The plasma ions located in the path of the positron beam result in a de-focussing force on it. The transport of the positron in a positron-beam driven wake is thus not ideal in a homogenous plasma and has to rely on external focusing optics ahead of the plasma. The use of hollow plasma channels with a few $c/\omega_{pe}$ is shown here to provide possible pathways to overcome these fundamental problems.

The formation of much shorter plasma channels excited by significantly different processes have been shown previously. These processes including using a collimated laser with annular profile \cite{milchberg-axicon-channel}\cite{bessel-beam-2011}, using a hollow capillary discharge \cite{hollow-channel-discharge}, among others \cite{ting-smlwfa-channel}\cite{ponderomotive-pdpwfa-wake-channel} \cite{ponderomotive-linear-wake-channel}. 


As the ion-wake channel is a practical realization of the hypothesized ideal hollow-channel plasma \cite{positron-accln-2001} of a few skin-depth channel radius, we examine its excitation by a positron-beam and possible use for positron acceleration \cite{positron-IPAC-2015}. In this section we analyze whether the positron-beam driven wake-fields excited in the ion-wake can be used for the acceleration and transport of a positron beam.

\subsection{Non-linearly driven Ion-wake channel: \\ analytical model }
\label{driven-ion-wake-analytical}

Positron acceleration using the ion-wake channel is explored in the non-linear ``crunch-in" regime of perturbed electron oscillation radii, $\delta r_e \geq r_{ch}$ under the condition that the peak beam density $n_{pb} > n_0$.

We use the analytical model of the radial electron ``crunch-in" based excitation of a positron beam wake in the plasma . The equation of motion of the plasma electron rings at $r$ from the axis, under the positron beam crunch-in force but neglecting the space-charge force of the collapsing electron rings is: $\frac{\operatorname d^2}{\operatorname d\xi^2}r \propto - \frac{1}{r} n_{bp}(\xi) r_{bp}^2(\xi)$, where $\xi = c\beta_{pb}t - z$ is the space just behind the positron beam with velocity $c\beta_{pb}$ driving the collapse. This is a non-linear second-order differential equation of the form, $r'' = f(r,r',\xi)$ where $f$ is not linear in $r$. Under the assumption about the positron-beam properties, $n_{bp}(\xi)$ and $r_{bp}(\xi)$ being constant during the entire interaction of the positron-beam with the hollow-channel over its full length. So, upon dropping the dependence on $\xi$ the equation simplifies to its {\it special case} which has analytical solutions, $r'' = f(r,r')$. The solution to this equation is \cite{positron-IPAC-2015}, $r_{ch}\sqrt{\pi} ~ \mathrm{erf} \left( \sqrt{\mathrm{ln}( r_{ch} /r)} \right) = -\sqrt{2\mathcal{C}} ~ \xi$,
where $\mathcal{C}=\frac{1}{2\pi\beta_b^2} \frac{n_{bp}}{n_0} \pi \left(\frac{r_{bp}}{c/\omega_{pe}}\right)^2$. Therefore, the collapse time-duration is $\xi_{coll} =  -r_{ch}\sqrt{\frac{\pi}{2\mathcal{C}}} $. We note that there is an anomaly that exists in our problem formulation and the solution because we have not taken into account the space-charge force of the compressing electrons as they collapse to the axis and this force balances the crunch-in force of the positron beam. Under these approximations the collapse time in a homogeneous plasma is \cite{positron-accln-2001}: 

\begin{align}
\tau_{c}=\sqrt{\pi} \frac{ r_{ch}}{\omega_{pe} \sqrt{ n_{bp}/n_0 } r_{pb} }
\label{positron-beam-electron-collapse}
\end{align} 

\noindent This expression shows that the collapse-time even in a homogeneous plasma depends strongly on the properties of the beam and the radius from which the rings are collapsing in.

Also, note that we have neglected the initial expansion velocity of the channel, $dr_{ch}/dt$). For optimal compression avoiding phase-mixing, the electron rings should collapse over, $\tau_c \simeq \mathcal{D}\lambda_{Np}/c$ where $\lambda_{Np}$ is the non-linear wavelength of the positron-driven wake and $\mathcal{D}$ is the duty-cycle of compression phase. So, the optimal channel radius is $r_{ch}^{opt} \simeq  2\sqrt{\pi}\mathcal{D}\frac{\lambda_{Np}}{\lambda_{pe}}\frac{\omega_{pb}}{\omega_{pe}} r_{pb}$. The scaling of the $r_{ch}^{opt}$ with positron beam parameters is shown in \cite{positron-IPAC-2015}.

\begin{figure}
	\begin{center}
   	\includegraphics[width=5in]{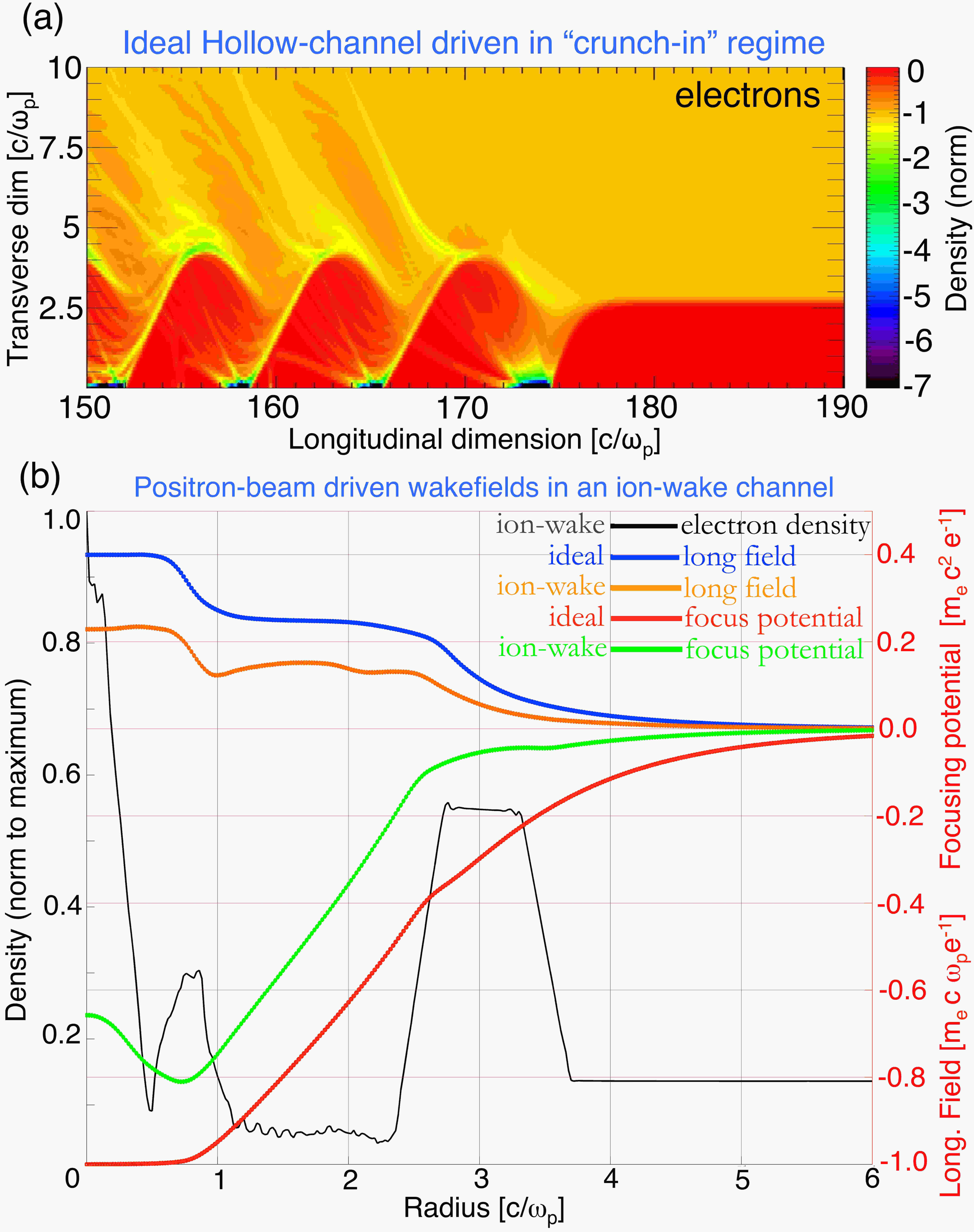}
	\end{center}
\caption{ \small {\bf 2D-PIC simulations (a) electron density in real-space showing electron density excitation in the ``crunch-in'' regime (b) corresponding radial profile of wakefields excited by a positron-beam in an ideal-channel versus an ion-wake channel}. The radial profile of the normalized electron density (black) in an ion-wake channel (normalized to the maximum electron compression) at longitudinal location of the peak accelerating wakefield ($r_{pb} = 2.3c/\omega_{pe}$, $\gamma_{pb}=38000$, $n_{pb}=1.3 n_0$). The radial profile of the accelerating-wakefield and normalized focussing-wakefield potential (radial field integrated from the edge of the box to a radius).}
\label{fig8:positron-acc-field-radial-profiles}
\end{figure}

\subsection{Non-linearly driven Ion-wake channel: \\ simulation result }
\label{driven-ion-wake-simulation}

Using 2-$\frac{1}{2}$D PIC simulations in a moving window we study the positron beam driven wakefields in cylindrical geometry. We compare positron acceleration in an ideal (Heaviside density function, $n_0\mathsf{H}(r-r_{ch})$) and an ion-wake channels (on-axis and channel-edge density-spike, channel minimum density of $0.1n_0$) with $r_{ch}= 2.5 ~ c/\omega_{pe}$. For non-linear wake parameters $r_{pb}=2.3 c/\omega_{pe}$, $n_{pb} = 1.3 n_0$ and $r_{ch}^{opt}\simeq 2.3 c/\omega_{pe}$ ($\mathcal{D}\frac{\lambda_{Np}}{\lambda_{pe}}=0.25$). Fig.\ref{fig8:positron-acc-field-radial-profiles} shows that the peak on-axis accelerating field is $0.4~m_ec\omega_{pe}e^{-1}$ for an ideal channel and $0.2~m_ec\omega_{pe}e^{-1}$ for the ion-wake channel. Fig.\ref{fig8:positron-acc-field-radial-profiles} also shows that the focussing potential (normalized to 27.6 $m_ec^2e^{-1}$) is similar and overall focussing in both cases. However, in the ion-channel the radial field is defocussing around the on-axis ion-spike. 

Thus the non-linearly driven ion-wake channel is useful for accelerating and transporting positrons despite the lower accelerating and focusing fields in comparison with the ideal channel. We also note that the on-axis density spike has detrimental effect on the focussing fields near the axis. However, the on-axis density spike is unstable over longer time-scales and collapses \cite{Sahai-dissertation}. The cylindrical simulations used in the current work to model the ion-wake ignore any azimuthal asymmetries in the distribution of electrons and ions in the on-axis density spike. Exploring the collapse of the on-axis density spike will be addressed in future-work. Ideal channels of a few $c/\omega_{pe}$ are technologically challenging whereas the ion-wake channel of radius $r_{ch}\gtrsim R_B$ is formed behind every bubble-wake.

\section{Conclusion}
In conclusion, using theory and PIC simulations we have shown the dynamics of the formation and evolution of a non-linear ion-wake excited by the well-characterized time-asymmetric electron bubble-wakefields independent of the type of energy-source. We have shown that the non-linear ion-wake has a characteristic cylindrical ion-soliton solution and evolves to an N-soliton solution over longer time as described by a driven cKdV equation. Thus over the period of persistence of the ion-soliton, a second electron bunch cannot be accelerated in the plasma. This establishes an upper limit on the repetition rate of a plasma collider. We have also shown the feasibility of using the ion-wake channel for positron acceleration in the positron-beam driven ``crunch-in" regime within an experimentally relevant parameter regime.

\begin{acknowledgments}
Work supported by the US Department of Energy under DE-SC0010012 and the National Science Foundation under NSF-PHY-0936278. I acknowledge the hospitality of the Dept. of Physics at the Imperial College London and the John Adams Institute, while making corrections to the manuscript. I acknowledge the OSIRIS code \cite{osiris-code-2002} for PIC simulations presented here. I acknowledge support for experiments by FACET group at Stanford Linear Accelerator Laboratory and Prof. M. Downer's group at University of Texas at Austin. I acknowledge the 256-node {\it Chanakya} server at Duke university.
\end{acknowledgments}

\appendix
\section{Considerations in the Non-linear ion-wake model}
\label{ion-wake-model-considerations}

There are several considerations and assumptions that underlie the non-linear ion-wake model. Here we briefly describe these and seek to differentiate the ion-wake from other phenomena. Primarily, we establish that the ion-wake is a collision-less phenomena and it is significantly different from diffusion. Secondly, as the ion-wake is formed behind the high phase-velocity non-linear electron plasma waves that are excited as the wakefields of near speed of light energy sources, it is significantly different from hole-boring which occurs in a plasma where the energy source has nearly zero group velocity.

We recognize that to study the time evolution of a wake-excited plasma, for establishing the duration over which it relaxes to  thermal equilibrium, both collisional and collision-less dynamics have to be considered along with the physics of recombination modes such as electron-ion recombination. However, in this work the dynamics of plasma is modeled under the collision-less approximation. Thus, diffusion is not important during the timescales over which the ion-wake is studied. We do not discuss recombination except mentioning that the ``afterglow" is dominated by volume recombination while localized effects cannot be ruled out.

In order to formally establish the difference between the density wave processes that occur over collision-less timescale in contrast to the ones that start dominating under collisions, we show the assumptions made to arrive at the dynamics of diffusion. The process of diffusion is modeled with a parabolic partial differential equation which is deduced from the ion-fluid equations under the assumption that the inertial response of the ions is much faster than the collisional timescales. 

The effect of collisions is introduced as a drag force, $m_in ~ \nu_{coll} ~ \langle \vec{v}_i \rangle$. The collisional drag force modifies the ion-fluid equation of motion as, $m_i n_i\frac{\operatorname d\vec{v}_i}{\operatorname dt} = mn\left(\frac{\partial \langle \vec{v}_i \rangle}{\partial t} + \langle \vec{v}_i \rangle \vec{\nabla} \cdot \langle \vec{v}_i \rangle \right) = \pm e ~ n\vec{E} -  \vec{\nabla}\mathcal{P}_e - m_i n_i ~ \nu_{coll} ~ \langle \vec{v}_i \rangle $ where $\nu_{coll}$ is the average electron-ion collision frequency and is obtained from the mean free path. Diffusion of plasma is thus driven by the charge-separation field, $\vec{E}$ and the thermal pressure, $\mathcal{P}_e$ while being impeded by the collisional drag. Upon ignoring the inertia of the ions, $\frac{\partial \vec{v}_i}{\partial t} = 0$, the equation for the ion velocity by diffusion in an isothermal plasma is, $\langle \vec{v}_i \rangle =  \pm \frac{e}{m ~ \nu_{coll}} ~ \vec{E} - \frac{k_BT_e}{m ~ \nu_{coll}} \frac{\vec{\nabla}n_i}{n_0}$. The characteristic parameter of diffusion is the diffusion coefficient or diffusivity $D = \frac{k_BT_e}{m ~ \nu_{coll}}$ and mobility $\mu = \frac{e}{m ~ \nu_{coll}}$ which depend upon the collision frequency. Using the gradient of the velocity in the continuity equation and ignoring the mobility $\mu$, leads to a Fick's law diffusion equation, $\nabla^2\frac{n_i}{n_0} \propto \frac{\partial }{\partial t} \frac{n_i}{n_0}$; characteristic of a parabolic equation. 

When the mobility is retained, the fluid equation is a moment of the Fokker-Planck equation which is the kinetic model of the collision-driven drift and diffusion. The diffusion equation thus cannot support wave-like solution because such solutions are characteristic of a hyperbolic partial differential equation. 

The solutions of linear and non-linear diffusion equations show the evolution of density profile by diffusion and can be obtained using the self-similar formulation. The self-similar solutions show the spatial and temporal evolution of the density to be exponentially decaying. In the non-linear case, the density can have a sharp-front as it decays. However, a soliton-like propagating solution cannot be described with diffusion equation. Hence, the cylindrical ion-soliton presented here is not diffusion but a wave phenomenon.

The electron bubble wake is excited by a sub-wavelength impulse of an ultra-short driver. In contrast, the ion-wake is excited as the ions undergo sustained interactions with the bubble fields within the spatial extent of the wake over several plasma electron oscillations. This happens because the electron wake-plasmon oscillations \cite{wakefield} have a near speed-of-light phase-velocity ($\beta_{\phi} \simeq 1$) but negligible group-velocity \cite{Vlasov} $\beta_g\approx 3 v_{th}^2/c^2$ (in the 1-D limit), where, $v_{th} \simeq \sqrt{k_BT_e/m_e}$ is the mean electron thermal velocity of the background plasma. Therefore a slowly-propagating train of coupled electron plasmons is excited in a cold collision-less plasma \cite{Vlasov}. A large difference between phase-velocity and the group velocity of the electron oscillations allows sustained field-ion interactions. It should be noted that high phase-velocity plasma electron waves are possible only in a cold plasma with appropriate density, $n_0$ that allows near speed-of-light propagation of the energy sources, $\beta_{es} \simeq 1 \approxeq \beta_{\phi}$. Ion-soliton modeled here is assuming a significant difference between the phase-velocity and the group velocity of the plasma-electron waves.

A time symmetric electron wakefield would excite time symmetric ion oscillations where the ion velocities average to zero. However, the bubble wake is asymmetric in time as the back of the bubble electron compression is a small fraction of the length of electron cavitation. The electron oscillations become non-linear at high driver intensities as all the interacting electrons are displaced radially, $\delta n_e/n_0 > 1$, forming a non-linear bubble-shaped electron spatial structure enclosing ions in its cavity. The wakefields excited in the bubble are useful for accelerating electrons \cite{cavitation-beam}\cite{cavitation-laser-expt}\cite{cavitation-beam-expt}. High intensities also lead to fields that can directly drive the plasma electrons to velocities near the speed-of-light. This occurs when for a laser pulse $a_0 \geq 1$ and an electron beam $\frac{n_b}{n_0}\left(\frac{r_b}{c/\omega_{pe}}\right)^2\geq 1$ where $a_0$ is the peak normalized laser vector potential, $n_b, r_b$ the peak beam density and radius. The radially expelled electrons oscillate radially under the force of the plasma ions. These oscillations are excited over plasma electron oscillation timescales, $2\pi\omega_{pe}^{-1}$ ($\omega_{pe}=\sqrt{4\pi n_0 e^2/\gamma_em_e}$) where $\gamma_e\beta_em_ec$ is the temporally anharmonic relativistic electron quiver momentum. The normalized quiver momentum of the electrons in the bubble-oscillations is relativistic $\gamma_{\perp}\beta_{\perp} \geq 1$ and the quiver frequency is $\omega_{\perp}=\omega_{pe} \left( \frac{\beta_{\phi}^2}{\gamma(1-\beta_{\phi}^2)} \right)^{1/2}$ \cite{Akhiezer-Polovin}. 

We show that non-linear ultra-relativistic electron wakefields interacting with the plasma ions lead to the excitation of a non-linear ion-wake. The non-linear ion-wake $\delta n_i/n_0 > 1$ in Fig.\ref{fig1:ion-wake-laser} and Fig.\ref{fig2:ion-wake-beam} is excited over timescales $\gg 2\pi\omega_{pe}^{-1}$ in the trail of a bubble-wake train. By shaping the energy source it can be matched or guided to excite a long train of nearly identical plasmons, Fig.\ref{fig3:bubble-train}. Since it is the electric field ${\bf E}_{wk}$ of a nearly stationary bubble plasmon that excite collective ion-motion we model the ion dynamics in a single bubble. Using the single bubble ion dynamics, Fig.\ref{fig4:bubble-ion-dynamics} we model the ion-wake over the whole bubble-train spanning several hundred plasma skin-depths ($c/\omega_{pe}$).

The wake-plasmon energy density ($\mathcal{E}_{wk} = 0.5 (e\lvert {\bf E}_{p}\rvert/(m_ec\omega_{pe}))^2 ~ m_e c^2 ~ n_0$, where ${\mathbf E}_{p}$ is the wakefield amplitude) is continually partitioned between the field energy and the coherent electron quiver kinetic energy. In our model we do not include heavy beam-loading of the bubble electron wake. Under heavy beam-loading the bubble field energy is efficiently coupled to the kinetic energy of the accelerated beam. In this scenario the bubble collapses and the magnitude of the ion-wake is smaller. The decoherence of the ordered electron quiver to random thermal energy, $\mathcal{E}_{wk} \rightarrow k_BT_{wk}$ due to the phase-mixing \cite{phase-mixing-longitudinal} of individual electron trajectories caused by the non-linearities and inhomogeneities is further stimulated by the ion motion. The details of the thermalization of the wake electrons under ion motion is beyond the scope of this paper. It is over these timescales upon thermalization that the steepened ion-density expands outwards radially as a non-linear ion-acoustic wave driven by the electron thermal pressure. The energy transfer process observed here is a coupling from the non-linear plasma electron-mode to a non-linear ion-acoustic mode \cite{lte-napac-2013}. We also observe energy coupling to the bow-shock which is formed behind the bubble, Fig.\ref{fig4:bubble-ion-dynamics}.

\newpage

\end{document}